\documentclass[aps,pra,twocolumn]{revtex4}
\usepackage[english]{babel}
\usepackage[dvips]{graphicx}
\usepackage{enumerate,amsthm,amsmath,amssymb,color,graphicx,bbm,float}
\usepackage{natbib}
\usepackage{hyperref}
\usepackage{epstopdf}

\def\d{{\rm d}}

\begin{document}

\title{Continuous-variable quantum key distribution protocols \\ with a non-Gaussian modulation}

\author{Anthony Leverrier}
\affiliation{ICFO-Institut de Cienc\`es Fot\`oniques, 08860 Castelldefels (Barcelona), Spain}

\author{Philippe Grangier}	
\affiliation{Laboratoire Charles Fabry, Institut d'Optique, CNRS, Univ. Paris-Sud,\\
Campus Polytechnique, RD 128, 91127 Palaiseau Cedex, France}

\date{\today}

\begin{abstract}
In this paper, we consider continuous-variable quantum key distribution (QKD) protocols which use non-Gaussian modulations. These specific modulation schemes are compatible with very efficient error-correction procedures, hence allowing the protocols to outperform previous protocols in terms of achievable range. In their simplest implementation, these protocols are secure for any linear quantum channels (hence against Gaussian attacks). We also show how the use of decoy states makes the protocols secure against arbitrary collective attacks, which implies their unconditional security in the asymptotic limit.
\end{abstract}
\maketitle

\section{Introduction}

The first potentially real-life application of the emerging field of quantum information is arguably quantum key distribution (QKD) which allows two distant parties to establish a secret key over an \emph{a priori} unsecure communication channel \cite{SBC08}. 
The importance of a QKD protocol is usually measured through three main criteria: its \emph{practicality}, its \emph{performance} and its \emph{security}.

Among all existing protocols, those based on continuous variables appear quite appealing from a \emph{practical} point of view \cite{CG07}. For instance, the protocol GG02 \cite{GG02} simply requires the preparation of coherent states and their detection via homodyne (or heterodyne) detection. 
Moreover, the \emph{security} of this protocol is established against collective attacks \cite{GC06,NGA06,LG10b} which are optimal in the asymptotic limit \cite{RC09}. 
There exist different ways to quantify the \emph{performance} of a given protocol, but the usual figure of merit is the secret key rate that can be achieved as a function of the distance. Continuous-variable protocols such as GG02 perform quite well for short distances, but seem unfortunately limited to much shorter distances than their discrete-variable counterparts (see Fig. 4 of Ref. \cite{SBC08} for a comparison of the performance of various protocols). In particular, while discrete-variable protocols allow one to distribute secret keys over distances larger than 100 km, GG02 has only be implemented for a distance of 25 km \cite{LBG07,FDD09} and there is not much hope of increasing its range well beyond 50 km \cite{LAB08}.

In this paper, we introduce continuous-variable (CV) QKD protocols which share the practicality and security of GG02, but greatly improve its performance, especially in terms of achievable distance. 

In Section \ref{philosophy}, we quickly review the status of current CV protocols. In Section \ref{new_schemes}, we present the specific modulation schemes of our protocols. We then introduce the concept of \emph{decoy states} for CV QKD protocols in Section \ref{section_decoy}.
In Section \ref{security}, we prove the security of our protocols against collective attacks and we finally discuss their performance in Section \ref{performance}.

\section{Limitations of current CV protocols}
\label{philosophy}

Proving the security of a QKD protocol is usually a difficult task, but the situation is even worse for continuous-variable QKD protocols, because the relevant Hilbert space is infinite dimensional. The task is difficult for (at least) one specific reason: the bipartite state $\rho_{AB}$ shared by Alice and Bob (in the entanglement-based version of the protocol) has an infinite number of degrees of freedom. This means that a full tomography of this state is hopeless, even in the case of collective attacks, where Alice and Bob share many copies of the same state. 

Fortunately, one can take advantage of extremality properties of Gaussian states \cite{WGC06} to show that the eavesdropper's information is upper-bounded by the information she would have if Alice and Bob shared instead the state $\rho_{AB}^G$, the Gaussian state with the same first two moments as $\rho_{AB}$ \cite{GC06,NGA06,LG10b}. Hence, knowing the first two moments of the state $\rho_{AB}$, that is, a \emph{finite} number of parameters, is sufficient to bound Eve's information. Furthermore, taking advantage of specific symmetries of the protocols in phase-space \cite{LKG09}, it is possible to reduce this number to only three, namely the variances of Alice and Bob's reduced states and the covariance.
That so few parameters are indeed sufficient to bound Eve's information is quite remarkable, but it should be noted that this is also a necessity if one wants to take finite-size effects into account. Indeed, it was shown in \cite{LGG10} that estimating these parameters with a precision compatible with a secret key rate is already very challenging in terms of resources.

Because of the constraints imposed by finite-size effects, the only theoretical tool presently available to prove the security of a continuous-variable QKD protocol is therefore this Gaussian optimality. This technique unfortunately comes at a price, namely that if the true quantum state $\rho_{AB}$ is not sufficiently close to a Gaussian state, then the bound on Eve's information will not be tight enough to still get secret bits at the end of the protocol. By construction, this bound is indeed only tight for Gaussian states, and it turns out that it degrades very rapidly as the non-Gaussianity of the state (and consequently of the protocol) increases.

This observation leads us to the unavoidable conclusion that with the theoretical techniques available today, protocols using a Gaussian modulation ({\it i.e.} GG02, possibly with a heterodyne detection \cite{WLB04}) are optimal among all continuous-variable QKD protocols, and using any other modulation scheme can only lead to worse performances \footnote{Note that restricting the eavesdropper to specific classes of attacks, such as beamsplitter attacks, allows one to compute tighter bounds on the secret key rate, even for non-Gaussian modulations, but this is not the case if one is not willing to arbitrarily restrict the possible attacks of the eavesdropper.}. 

A natural question then arises: why should one consider new protocols involving specific non-Gaussian modulations since \emph{theoretically}, they cannot beat a Gaussian modulation?
The reason is that, \emph{in practice}, protocols with a Gaussian modulation do not perform as well as expected, especially with regards to long distance communication. This is due to the fact that error-correction is very hard to implement for a Gaussian modulation, thereby making the effective secret key rate drop to zero at distances of the order of 50 km, even if the theoretical key rate is strictly positive.

The protocols we introduce in the present paper outperform previous protocols in terms of maximal range. This is achieved thanks to a combination of two facts: first, the specific modulation schemes allow one to extract information much more efficiently than with a Gaussian modulation ; second, there is a regime (for the variance of the modulation) where the protocols are still significantly close to a Gaussian protocol, hence making the bound on Eve's information tight enough to be useful.

\section{New modulation schemes}
\label{new_schemes}

The main argument for switching from a Gaussian modulation to a non-Gaussian one is \emph{not} because it might make the eavesdropping more difficult (available theoretical tools are not powerful enough to answer this question). The reason is that present coding techniques do not allow for a very efficient reconciliation procedure at (very) low Signal-to-Noise Ratio (SNR) when Alice's data are Gaussian. Remember that a QKD protocol typically consists of three phases: first, Alice and Bob exchange quantum signal, perform measurements and obtain correlated classical data, say vectors {\bf x} and {\bf y}; second, in the \emph{reconciliation} step, they use classical error correction techniques to agree on a common (errorless) bit string {\bf u}, and finally, they apply \emph{privacy amplification} to obtain a secure key from {\bf u}. 

The importance of the reconciliation phase is specific to continuous-variable protocols for two reasons. First, one has to deal with continuous values for \emph{both} Alice and Bob's variables instead of bits, which is rather unusual in the field of digital communication. Indeed, even when analog signals are used to transmit information on noisy (classical) channels, the modulation is almost always discrete and not continuous. The second reason is that contrary to discrete-variable QKD protocols where the error rate is always below some small constant (to guarantee the security of the protocol), the SNR can be arbitrarily small for continuous-variable protocols (and is actually very small as one tries to increase the range of the protocol \footnote{As a first order estimate, if the variance of modulation is $V_A$ and the transmission of the channel is $T \approx 10^{-0.02 d}$ (in fiber if $d$ is the channel length in kilometers), then the SNR is roughly given by $T V_A$. The modulation variance on Alice's side, $V_A$ is a free parameter of the protocol which can be optimized for a quantum channel, but numerical simulations show that the optimal value for $V_A$ practically does not depend on the transmission $T$ of the channel. For this reason, the SNR is roughly proportional to the transmission of the channel, and is consequently very small for long distances.}).

For continuous-variable QKD protocols, information is encoded in phase space, in general in the quadratures of coherent states \footnote{In principle, Alice could alternatively encode information on squeezed states, but this is not practical and does not lead to vastly superior performances.}. More precisely, if the classical information she wants to send is described by a vector ${\bf x} = (x_1,x_2,\cdots, x_{2n})$, then Alice prepares the $n$ coherent states $|x_1+ i x_2\rangle, \cdots, |x_{2n-1} + ix_{2n}\rangle$ and sends them through the quantum channel. 

There are two possibilities concerning the detection: Bob can perform either homodyne or heterodyne measurements. In the case of a homodyne detection, for each state, Bob chooses randomly which quadrature to measure. He then obtains an $n$-dimensional classical vector ${\bf y}$ and later informs Alice of his choices of quadratures. In the case of a heterodyne detection, Bob ends up with a $2n$-dimensional vector ${\bf y}$. Hence, Alice and Bob share twice as many data for the heterodyne protocol than for the homodyne one. However, a heterodyne detection adds 3 dB of noise to the data, and in practice the performances of both schemes are quite similar.

In order to complete the key distribution, two additional steps are required. First Alice and Bob proceed with the reconciliation of their classical data in order to agree on a common bit string $u$. Here, we restrict ourselves to a \emph{reverse} reconciliation \cite{GVW03}, meaning that only Bob can send classical information to Alice (in contrast to \emph{direct} reconciliation where Alice would send some side-information to help Bob correct his errors). Second, they use privacy amplification in order to obtain a secret key from $u$: this can only be done once they have an upper bound on Eve's information about $u$. 

In this paper, we consider protocols for which the reconciliation can be performed efficiently (which is not the case of the protocols using a Gaussian modulation), and for which Eve's information can be bounded if she is restricted to collective attacks. Security against general attacks (``unconditional security")  is then obtained in the asymptotic limit, thanks to a de Finetti representation theorem for infinite-dimensional quantum systems \cite{RC09}.

To be more specific, the protocols considered here are characterized as the ones for which the \emph{reverse} reconciliation problem can be reduced to the channel coding problem for the binary-input additive white Gaussian noise (BI-AWGN) channel, which can itself be tackled with standard techniques (efficient error correcting codes). One such protocol is the four-state protocol considered in \cite{LG09}, but it turns out that other, more efficient, \emph{continuous} modulation schemes are also possible.

\subsection{Homodyne detection~: four-state modulation}

Let us consider protocols involving a homodyne detection. In this case, the protocols of interest display a discrete modulation with either 2 or 4 states, and a basic problem is how to evaluate the transmission channel, and thereby Eve's information. Such  a problem does not arise for Gaussian modulations, because the variances and covariances directly measured by Alice and Bob give a covariance matrix, which is all that is needed to characterize the worst possible attack by Eve, which is a Gaussian attack according to the Gaussian optimality theorem. On the other hand, for a non-Gaussian modulation, the noises and correlations  measured by Alice and Bob cannot be directly connected to a relevant covariance matrix, and the Gaussian optimality theorem does not apply directly. 

Among possible approaches, two-state protocols were studied in \cite{ZHR09}, and the authors proved the security of the protocols against any collective attacks, but with the \emph{caveat} that a complete tomography of the state was required. In \cite{LG09,LG10}, the authors considered the noises and correlations  measured by Alice and Bob from their non-Gaussian modulation, and then considered the Gaussian attacks associated  with these values as optimal. However, this approach has the implicit assumption that the transmission channel can be considered as \emph{linear}, which means that it is intrinsically  characterized by a transmission $T$ and an excess noise $\xi$ (see details in Appendix \ref{linear}). Gaussian channels are obvious examples of linear channels, but a linear channel may also be non-Gaussian.  
Since the proofs of \cite{LG09,LG10} only involve estimating the second moments of Alice and Bob's correlations, they are compatible with a practical implementation taking into account finite-size effects \cite{LGG10}, but  they are not fully general.

In the present paper, we extend the security proof of discrete modulation protocols against arbitrary collective attacks. Our proof still only requires us to estimate two quantities: the transmission $T$ and excess noise $\xi$ of the quantum channel, without carrying out a full channel tomography. But achieving this estimation requires  the use of decoy states,  as will be explained in Section \ref{section_decoy}.

Since the four-state protocol always outperforms the two-state protocol, we will only discuss the case of the former in the present paper. 
In this protocol, the modulation consists of four coherent states: $|\alpha e^{i \pi/4}\rangle, |\alpha e^{3i \pi/4}\rangle, |\alpha e^{-3i \pi/4}\rangle, |\alpha e^{-i \pi/4}\rangle $ where $\alpha$ is a positive number. The modulation variance $V_A$ is given by $V_A=2\alpha^2$.

The practical implementation of the reverse reconciliation problem for the four-state protocol is discussed in details in Appendix \ref{reduction}.

\subsection{Heterodyne detection: non-Gaussian continuous modulations}

Before introducing the modulation schemes compatible with a heterodyne detection, let us say a few words concerning the reconciliation procedure. The main difficulty here lies in the fact that we need a reverse reconciliation. Indeed, the side information sent by Bob must help Alice without giving Eve any relevant information. 
The only schemes where side information seems to have these properties (while being efficiently computable by Bob) are when it describes rotations in particular dimensions, namely dimensions 1, 2, 4 and 8 \cite{LAB08}. This surprising result is a consequence of the fact that the only real division algebras are the real numbers ($\mathbbm{R}$), the complex numbers ($\mathbbm{C} \cong \mathbbm{R}^2$), the quaternions ($\mathbbm{H} \cong \mathbbm{R}^4$) and the octonions ($\mathbbm{O} \cong \mathbbm{R}^8$). Indeed, one can show that the possibility of an efficient reconciliation protocol (in terms of computation complexity as well as classical communication) is intimately connected with the existence of a division operation for the data considered. In particular, identifying vectors in $\mathbbm{R}$, $\mathbbm{R}^2$, $\mathbbm{R}^4$ or $\mathbbm{R}^8$ with units of the real numbers, the complex numbers, the quaternions or the octonions allows one to take benefit of the division structure naturally associated with these ensembles.
For instance, the reconciliation protocol of the four-state protocol exploits this property in dimension 1. The modulation schemes we consider now exploit it for dimensions 2, 4 and 8.

First, note that in the case of the four-state protocol, Alice chooses the value of each quadrature uniformly on the (0-dimensional) sphere $\mathcal{S}^0 = \{-1, 1\}$ in dimension 1. (This value is then appropriately rescaled to obtain the desired variance of modulation.) For this reason, we will sometimes refer to the four-state protocol as the 1-dimensional protocol in the rest of the paper.

The modulation schemes we consider now are simply the generalizations to dimensions 2, 4 and 8. Hence, in the $d$-dimensional protocol (with $d \in \{2,4,8 \}$), Alice draws random variables uniformly on the sphere $\mathcal{S}^{d-1}$ in dimension $d$. 

For $d=2$, Alice draws $n$ points on the unit circle: $\{ (x_1^1, x_1^2),  (x_2^1, x_2^2), \cdots,  (x_n^1, x_n^2) \}$, and sends the $n$ coherent states $|x_1^1 + i x_1^2\rangle, \cdots, |x_n^1 + i x_n^2\rangle$, where the variables are rescaled by a factor $\alpha$.

For $d=4$, Alice draws $n/2$ points on the sphere $\mathcal{S}^3$: $\{ (x_1^1, x_1^2, x_1^3, x_1^4), \cdots,  (x_{n/2}^1, x_{n/2}^2, x_{n/2}^3, x_{n/2}^4) \}$, and sends the $n$ coherent states $|x_1^1 + i x_1^2\rangle, |x_1^3 + i x_1^4\rangle \cdots, | x_{n/2}^3 + i  x_{n/2}^4\rangle$, where the variables are rescaled by a factor $\alpha \sqrt{2}$.

For $d=8$, Alice draws $n/4$ points on the sphere $\mathcal{S}^7$: $\{ (x_1^1, x_1^2, x_1^3, x_1^4, x_1^5, x_1^6, x_1^7, x_1^8), \cdots,  (x_{n/4}^1, x_{n/4}^2, x_{n/4}^3, x_{n/4}^4$,  $x_{n/4}^5, x_{n/4}^6, x_{n/4}^7, x_{n/4}^8) \}$, and sends the $n$ coherent states $|x_1^1 + i x_1^2\rangle, |x_1^3 + i x_1^4\rangle \cdots, |x_{n/4}^7 + i  x_{n/4}^8\rangle $, where the variables are rescaled by a factor $2\alpha$.

The procedure to reduce the reverse reconciliation problem in these three scenarios to the usual problem of channel coding for the BI-AWGN channel is explained in detail in Appendix \ref{reduction}.

Let us say a few words about what we mean by \emph{efficient} reconciliation procedure in the context of QKD. Usually in the field of computer science, an algorithm is said to be efficient if the amount of resources (\emph{e.g.} running time, randomness generation, classical communication, etc) it consumes grow at most polynomially with the natural size of the problem. In the case of a reconciliation procedure, the problem size is given by the length $n$ of the vectors Alice and Bob try to agree on. As $n$ usually takes very large values (for instance $10^{10}$ or $10^{12}$), an algorithm requiring resources scaling as $n^2$ or $n^3$ is obviously unacceptable: only a linear scaling is compatible with a practical implementation. Moreover, the factor of proportionality should be small enough. For $d = 1, 2, 4$ or 8, the reconciliation procedure introduced in \cite{LAB08} requires that Bob draws one random bit and transmits classically one real number to Alice  per exchanged signal. The computational complexity of the protocol (not including the decoding of the error correcting code) is linear with $n$. For values of $d$ strictly greater than 8, the naive approach (which consists in drawing random transformations uniformly in the orthogonal group $O_d$ and transmitting them to Alice) requires that Bob draw $d$ random variables from a normal distribution (instead of only 1 bit) and to send $d$ real values to Alice for each exchanged quantum signal, which is prohibitive for a realistic implementation. 

It should be emphasized that the higher the dimension, the higher the secret key rate of the QKD protocol. The reason for this is the very specific technique that we use to bound Eve's information. Our bound depends only on the covariance matrix of Alice and Bob's bipartite state in the entanglement-based version of the protocol, and therefore is tight only when the state is Gaussian. Fortunately, if the state is almost Gaussian, then the bound is good enough for our purpose. Because of this, we want to use a protocol as Gaussian as possible. It turns out that considering modulations in higher and higher dimensions brings us closer and closer to the Gaussian modulation for which the bound is tight. Indeed, a Gaussian modulation of variance 1 can be seen as drawing uniformly a random point of the sphere of radius $\sqrt{d}$ in $\mathbb{R}^d$ as $d$ tends to infinity.
Hence, the GG02 protocol with a Gaussian modulation can be seen as the $d$-dimensional protocol with $d=\infty$. Unfortunately, for $d=\infty$, efficient reconciliation techniques at low SNR are not known.

\section{Decoy states}
\label{section_decoy}

As we already mentioned, it is crucial that the security proof of CV QKD protocols  requires the estimation of only a few parameters. Ideally, one would like a GG02-type security proof where only the transmission $T$ and the excess noise $\xi$ of the quantum channel need to be estimated.

As we will discuss in Section \ref{security}, the security proof of our protocols indeed relies on the fact that one can estimate the covariance matrix of Alice and Bob's bipartite state in the entanglement-based version of the protocol. 

A difficulty lies in the fact that Alice and Bob do not perform the entanglement-based version of the protocol (in which case the covariance matrix is directly accessible in the experiment) but use instead the \emph{prepare and measure} version. Hence, if Alice encodes the variable $x$ in the quadrature of a state and Bob obtains the result $y$ when measuring this quadrature, they can estimate the three following moments of order 2: Alice's variance $\langle x^2 \rangle$, Bob's variance $\langle y^2 \rangle$ and the covariance $\langle x y \rangle$. Whereas Alice and Bob's variance in the prepare and measure scenario are directly related to the respective variances in the entanglement-based scenario, the same is not true for the covariances. 

There are two cases where the covariance matrix in the prepare and measure protocol allows one to recover the covariance matrix of the state in the entanglement-based scenario, namely when the quantum channel is linear (see Appendix \ref{linear}), for instance a Gaussian channel, and when the modulation is Gaussian. 

In Refs. \cite{LG10, LG10c}, the security of the protocols considered in the present paper was established in the case of linear channels. Here, we wish to get rid of this hypothesis (which can never be perfectly verified in practice with a finite number of samples), and for this reason, it is necessary to use a Gaussian modulation for the parameter estimation procedure. Unfortunately, it is not \emph{a priori} possible to use two different modulations for key distribution and parameter estimation, because an eavesdropper could use a different strategy in each case. The solution is to add a \emph{third} modulation consisting of decoy states. Let us call ``key'', ``decoy'' and ``G'' the modulations corresponding respectively to states used for the key distillation, decoy states and states used for parameter estimation purposes (a Gaussian, in fact thermal, distribution). One can define the three following states:
\begin{eqnarray}
\sigma_{\mathrm{key}}^d &=& \int p_\mathrm{key}(\alpha) \, |\alpha\rangle\!\langle \alpha |  \, \mathrm{d} \alpha\\
\sigma_{\mathrm{decoy}}^d &=& \int p_\mathrm{decoy}(\alpha) \, |\alpha\rangle\!\langle \alpha |  \, \mathrm{d} \alpha \label{decoy_def} \\
\sigma_\mathrm{G}^d &=& \int p_\mathrm{G}(\alpha) \, |\alpha\rangle\!\langle \alpha | \, \mathrm{d} \alpha 
\end{eqnarray}
where $\alpha \in \mathbbm{R}^d$ for the $d$-dimensional protocol. (In particular, $|\alpha\rangle$ refers here to $d/2$ coherent states.) In these expressions, the probability distribution $ p_\mathrm{key}$ is the uniform measure of the sphere $\mathcal{S}^{d-1}$ (with radius $\alpha \sqrt{d/2}$) and $p_\mathrm{G}$ is the Gaussian distribution $\mathcal{N}(0,\alpha^2/2)^{\otimes d}$ in $d$ dimensions. In other words, $ p_\mathrm{key}$ corresponds to the modulation schemes described in Section \ref{new_schemes}, and $p_\mathrm{G}$ is the Gaussian distribution of the GG02 protocol.

If the probability distribution $ p_\mathrm{decoy}$ is chosen such that 
\begin{equation}
\label{decoy}
p \, \sigma_{\mathrm{key}}^d + (1-p)\, \sigma_{\mathrm{decoy}}^d = \sigma_\mathrm{G}^d,
\end{equation}
where $p$ is a weight between 0 and 1, then the state sent by Alice to use for parameter estimation is indistinguishable from that used to distill a key (or as a decoy). 
The idea is that after the exchange of quantum states is complete, Alice can announce to Bob which states he can use for the key, which states he can discard (decoys) and which states should be used for parameter estimation. 

If in principle, the form of $\sigma_{\mathrm{decoy}}^d$ does not matter (it could be any state, not necessarily of the form \ref{decoy_def}), this is no longer true if we require the protocol to be practical. Indeed, for this reason, we impose the extra constraint that $\sigma_{\mathrm{decoy}}^d$ should be obtained as a mixture of coherent states, \emph{i.e.} be of the form \ref{decoy_def}. We discuss this in details in Appendix \ref{decoys_coherent} where we describe two techniques for finding the appropriate decoy states.

To summarize, the modulation that is used in the protocols is a mixture of three particular modulations.
Let us note $p_\mathrm{est}$ the fraction of states that Alice and Bob want to use for parameter estimation purposes (this fraction can be optimized numerically but in a typical scenario, its value can be around 50 $\%$).
Then for each state she sends, Alice will choose either the modulation $p_\mathrm{key}(\alpha)$ with probability $p(1-p_\mathrm{est} )$ or modulation $p_\mathrm{G}(\alpha)$ with probability $p_\mathrm{est}$ or send a decoy state with probability $(1-p)(1-p_\mathrm{est} )$.

\section{Security against collective attacks}
\label{security}

In this paper, we restrict ourselves to the case of collective attacks since they are optimal in the asymptotic limit \cite{RC09}. The (asymptotic) secret key rate $K$ is then given by \cite{DW05}:
\begin{equation}
\label{DW}
K = \beta I(A;B) - \chi (B;E),
\end{equation}
where $\beta$ is the reconciliation efficiency, $I(A;B)$ is the classical mutual information between Alice and Bob's data (for the data corresponding to the modulation $p_\mathrm{key}(\alpha)$) and $\chi(B;E)$ is the Holevo quantity:
\begin{equation}
\label{Holevo}
\chi(B;E) = S(\rho_E) - \sum_y p(y) S(\rho_{E|y}),
\end{equation}
where $S$ is the von Neumann entropy, $y$ is Bob's measurement result obtained with probability $p(y)$, $\rho_{E|y}$ is the corresponding state of Eve's ancilla and $\rho_E = \sum_y p(y) \rho_{E|y}$ is Eve's partial state.

Note that this rate should be modified to take finite-size effects into account. For simplicity, we only consider the asymptotic rate here, but a complete analysis of finite-size effects can be found in Ref. \cite{LGG10}.

Since the quantity $\beta I(A;B)$ is directly observable in practice, the goal of the security proof consists in deriving an upper bound for the quantity $\chi(B;E)$ which should be a function of parameters accessible in an experiment. In our case, we will find a bound which only depends on three parameters: the variance of modulation $V_A$ which is chosen by Alice, the transmission $T$ and the excess noise $\xi$ of the quantum channel which can be estimated with the technique described in Ref. \cite{LGG10}.

We now consider the entanglement-based version of our protocols.
In this scenario, Alice prepares $n$ two-mode squeezed vacuum states, keeps one half of each state and sends the second half to Bob through the quantum channel.

Let us introduce some notations. In the following, we will consider bipartite states, either before or after the quantum channel. We use the superscript 0 to denote states before the quantum channel. Moreover, the action of the quantum channel can be described by a map $\mathbbm{1}\otimes \mathcal{T}$ where the identity acts on the first part of the state (namely, Alice's state) and the quantum channel $\mathcal{T}$ acts non-trivially only on the second part of the state.

The three states of interest are $\rho_{G}^0$, $\rho_{\mathrm{key}}^0$ and $\rho_{\mathrm{decoy}}^0$, which are the  Schmidt purifications of the states $\sigma_{G}^d$, $\sigma_{\mathrm{key}}^d$ and $\sigma_{\mathrm{decoy}}^d$, respectively. (Note for instance that $\rho_{G}^0$ is a two-mode squeezed vacuum: $\rho_{G}^0 = |\mathrm{EPR}\rangle \!\langle \mathrm{EPR}|$.) After the quantum channel, these three states become respectively $\rho_{G}$, $\rho_{\mathrm{key}}$ and $\rho_{\mathrm{decoy}}$.

The main idea of the security proof is that one can bound Eve's information by a function of the covariance matrix of the state used to distill the key: $\rho_\mathrm{key}$. In the protocol, Alice always starts with a two-mode squeezed state $\rho_{G}^0$ but she can choose between two measurement strategies depending on whether a given state will be used for key distillation or for parameter estimation.

We also introduce a general measurement acting on Alice's part $\{\Pi_d, \mathbbm{1}-\Pi_d\}$ such that, when applied to the two-mode squeezed vacuum $\rho_{G}^0$, the result corresponding to $\Pi_d$ prepares the state $\rho_{\mathrm{key}}^0$ used in the $d$-dimensional protocol while the second result prepares $\rho_{\mathrm{decoy}}^0$ (see Appendix \ref{measurement} for a description of this measurement). 

For each state, Alice chooses randomly between key distillation and parameter estimation. The fraction of each task should be optimized taking into account all finite-size effects. 

If a state is dedicated to parameter estimation, Alice simply performs a heterodyne detection on her part and Bob proceeds as usual. At the end of the protocol, Alice and Bob can compare their statistics and compute the covariance matrix $\Gamma_G$ of the state $\rho_{G}$. For this, we do not need to make any assumption (for instance of linearity) concerning the quantum channel.

If a state is to be used for key distillation, then Alice performs the generalized measurement $\{\Pi_d, \mathbbm{1}-\Pi_d\}$ on her half of the state, thus preparing $\rho_{\mathrm{key}}^0$ with probability $p$ and $\rho_{\mathrm{decoy}}^0$ with probability $1-p$. States corresponding to the result $\mathbbm{1}-\Pi_d$ will later be discarded. Finally, only the states $\rho_{\mathrm{key}}^0$ are used for key distillation. Let us note $\Gamma_\mathrm{key}^0$ ($\Gamma_\mathrm{key}$) the covariance matrix of these states before (after) the quantum channel.

It was proven in \cite{GC06} that the quantity $\chi(B;E)$ can always be upper bounded by a function of the covariance matrix of Alice and Bob's bipartite state. Here, the state used for key distillation is $\rho_{\mathrm{key}}$, meaning that one can bound $\chi(B;E)$ with a (known) function of $\Gamma_{\mathrm{key}}$ (see Ref. \cite{LBG07} for the precise form of this function).

All that is left to do is therefore to compute the covariance matrix $\Gamma_{\mathrm{key}}$. Note that the covariance matrix $\Gamma_{\mathrm{key}}^0$ of the state before the quantum channel can be computed and only depends on the modulation variance (as well as the dimension of the modulation scheme). Details on how to compute this covariance matrix are given in Appendix \ref{covariance}.

\begin{figure}[th]
    \includegraphics[width=0.95\linewidth]{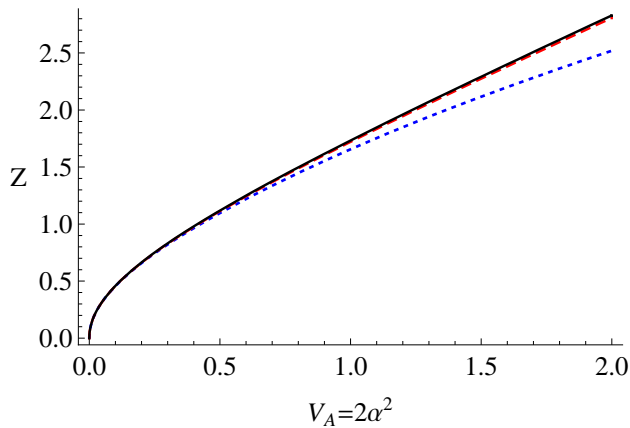}
    \caption{\label{Z}(Color online) Comparison of the covariance coefficient $Z$ for the states $\rho^0_\mathrm{EPR}$ (top solid black curve) and $\rho_\mathrm{key}^0$ for the four-state protocol (bottom short-dashed blue curve) and the 8-dimensional protocol (middle long-dash red curve) as a function of the variance of modulation $V_A$.}
\end{figure}

The covariance matrix $\Gamma_{\mathrm{key}}^0$ has the following form (with the convention $x_A, p_A, x_B, p_B$):
\begin{equation}
\Gamma_{\mathrm{key}}^0 \!= \!\left(\!
\begin{matrix} 
(V_A + 1) \mathbbm{1}_{\! 2} & \,Z_d\, \sigma_z  \\
\,Z_d \, \sigma_z & (V_A + 1)  \mathbbm{1}_{\! 2} \\ 
\end{matrix}\! \right)
\end{equation}
where $\sigma_z=\mathrm{diag}(1,-1)$.
Here $V_A = 2\alpha^2$ is Alice's modulation variance in the prepare and measure version of the protocol and $Z_d$ is a function of $V_A$ and the dimension $d$ of the protocol. The two-mode squeezed vacuum, corresponding to a modulation on a sphere whose dimension tends to infinity, has the same form with $Z_\infty := Z_\mathrm{EPR}= \sqrt{V_A^2+2V_A}$. A comparison of $Z_1$ (four-state protocol), $Z_8$ with the maximal value $Z_\mathrm{EPR}$ is displayed on Figure \ref{Z}.

One also knows the covariance matrix $\Gamma_G$ of the state $\rho_{G}$ \emph{after} the quantum channel:
\begin{equation}
\Gamma_G \!= \!\left(\!
\begin{matrix} 
(V_A + 1) \mathbbm{1}_{\! 2} & \sqrt{T} \,Z_\mathrm{EPR}\, \sigma_z  \\
\sqrt{T} \,Z_\mathrm{EPR} \, \sigma_z & (1 + T V_A + T \xi)  \mathbbm{1}_{\! 2} \\ 
\end{matrix}\! \right)
\end{equation}
where $T$ and $\xi$ refer respectively to the transmission and excess noise of the channel and can be estimated experimentally \cite{GCW03}.

The last part of the argument consists in proving that the covariance matrix  $\Gamma_{\mathrm{key}}$ has the same form (or at least can be safely considered to have the same form) as $\Gamma_G$ after the quantum channel, if one simply replaces $Z_\mathrm{EPR}$ by $Z_d$:
\begin{equation}
\label{Gamma_key}
\Gamma_{\mathrm{key}} \!= \!\left(\!
\begin{matrix} 
(V_A + 1) \mathbbm{1}_{\! 2} & \sqrt{T} \,Z_d\, \sigma_z  \\
\sqrt{T} \,Z_d \, \sigma_z & (1 + T V_A + T \xi)  \mathbbm{1}_{\! 2} \\ 
\end{matrix}\! \right)
\end{equation}

This is clear if the quantum channel $\mathcal{T}$ is linear, for instance Gaussian but the argument is more involved in the case of an arbitrary quantum channel.

In fact, if the channel is linear, Alice and Bob can directly compute the covariance matrix $\Gamma_{\mathrm{key}}$ from the data corresponding to the modulation $p_\mathrm{key}(\alpha)$ and the Gaussian modulation or the decoy states are not required in that case \cite{LG10,LG10c}.

Because the modulation considered here is indistinguishable from a Gaussian modulation, the parameter estimation is performed in such a way that there are no privileged direction in phase space.
However, this is formally not enough to warrant that Eve's attack has the same symmetry. We therefore provide the following two strategies: either one assumes that the symmetry is not broken by Eve (which in theory could be checked by performing a tomography of the state), or one does not want to make such an assumption and prefers to actively symmetrize the protocol.

In the first scenario, under the assumption that the symmetry of the quantum channel is not broken, the security protocol with decoy states presented here is at least as good as in the case where the channel is linear. Indeed the quantum state shared by Alice and Bob is invariant under the group of conjugate passive symplectic operations applied on Alice's $n$ modes and Bob's $n$ modes, which means that their state can be safely considered to be Gaussian if the analysis is restricted to collective attacks \cite{LG10b}.

However, if one does not want to rely of the assumption that the symmetry is not broken, it is possible to remove this assumption thanks to an active symmetrization of the protocol. This is described in detail in Appendixes \ref{active} and \ref{symmetrized_protocol}. This additional step shows that the state $\rho_G$ after the quantum channel is rotationally invariant in phase space. Therefore, when restricting ourselves to collective attacks, we conclude using the technique presented in Ref. \cite{LG10b} that the state  $\rho_G$ can be safely considered to be Gaussian. In particular, the covariance matrix given in Eq. \ref{Gamma_key}  can be used for the security analysis, with the same values of $T$ and $\xi$ as the ones in $\Gamma_G$, obtained from the parameter estimation step.

Finally, using the covariance matrix $\Gamma_{\mathrm{key}}$, one can compute the quantity $\chi(B;E)$ using for instance the formalism detailed in \cite{LBG07}.

\section{Performance of the protocols}
\label{performance}

The (asymptotic) secret key rate of the protocols reads:
\begin{equation}
K = \beta I(A;B) - \chi(B;E).
\end{equation}
The idea in our protocols is to use a modulation scheme which is compatible with a very efficient reconciliation, thereby greatly increasing the quantity $\beta I(A;B)$ in comparison with Gaussian modulation protocols. 

\begin{figure}[th]
    \includegraphics[width=0.95\linewidth]{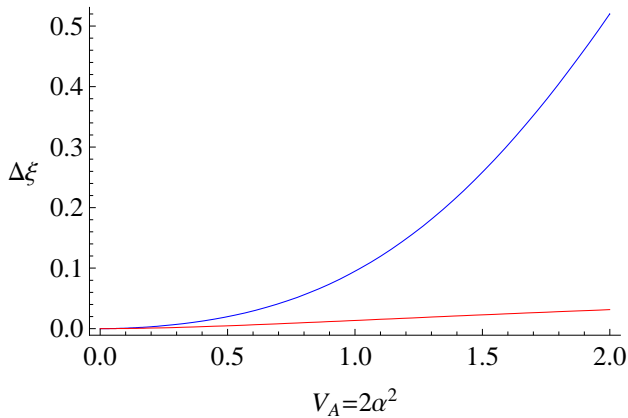}
\caption{\label{DeltaXi}(Color online) Equivalent excess noise $\Delta \xi$ due to the non-Gaussian modulation as a function of the variance of modulation $V_A$. Upper curve refers to the four-state protocol, lower curve to the 8-dimensional protocol. By definition, a protocol with a Gaussian modulation does not display any equivalent excess noise. An excess noise of one unit of shot noise corresponds to an entanglement-breaking channel, therefore no security is possible with such a level of noise. This figure clearly shows that the 8-dimensional protocol outperforms significantly the four-state protocol.}
\end{figure}

The price to pay is that this non Gaussianity makes our bound on $\chi(B;E)$ less tight. This is because the correlation $Z_d$ of the state $\rho_\mathrm{key}^0$ is strictly less than $Z_\mathrm{EPR}$ for a given variance of modulation. Interestingly, this discrepancy can be interpreted in terms of excess noise: the fact that $\rho_\mathrm{key}^0$ displays smaller correlations than the two-mode squeezed state has the same effect as some virtual excess noise. 
In particular, the value of $\chi(B;E)$ one obtains in the $d$-dimensional protocol corresponds to the value one would obtain for a Gaussian modulation protocol (GG02) with a quantum channel characterized by a transmission $T_G = T/F \approx T$, and  an excess noise $\xi_G = F \xi + (F-1) V_A \approx \xi + (F-1) V_A$, where $F \equiv (Z_{\text{EPR}}/Z_d)^2$.
Since one has $F \approx 1$ for reasonable values of $V_A$ (see Figure \ref{Z}), the main effect of the non-Gaussian modulation is the \emph{equivalent excess noise} $\Delta \xi = (F-1) V_A$. Figure \ref{DeltaXi} displays this equivalent excess noise in the case of the four-state protocol ($d=1$) and the 8-dimensional protocol.

In state-of-the-art implementations \cite{LBG07,FDD09}, the excess noise is typically less than a few percent of the shot noise. This gives an approximate limit for the value of the equivalent excess noise that is acceptable. In particular, for the four-state protocol, one needs to work with modulation variances below 0.5 units of shot noise. On the contrary, it becomes possible to work with much higher variances in the case of 8-dimensional protocol. 

\begin{figure}[th]
  \centerline{
    \includegraphics[width=0.95\linewidth]{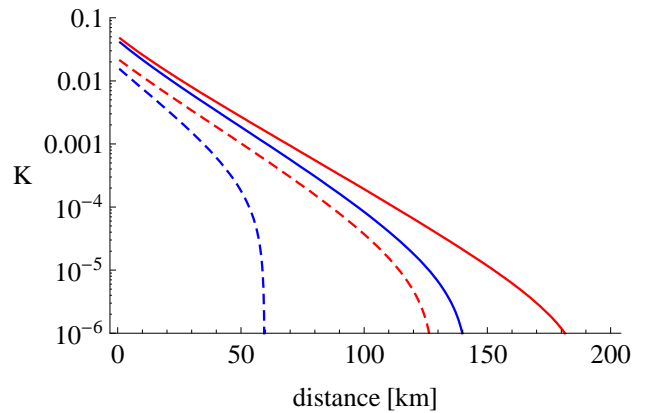}}
 \caption{\label{K4octo_asympt}(Color online) Asymptotic secret key rate $K$ for the 8-dimensional protocol (solid lines) and the four-state protocol (dashed lines) as a function of the distance (assuming transmission through a standard telecommunications fiber with 0.2 dB of loss per kilometer). The various parameters are an excess noise of 0.005 (upper red lines) or 0.01 (lower blue lines) and a quantum efficiency of the detectors $\eta$ of $60\%$. Reconciliation efficiency is supposed to be a conservative $80\%$.}
\end{figure}

In Fig. \ref{K4octo_asympt}, we display the asymptotic secret key rate of the four-state and the 8-dimensional protocols. The various parameters are chosen conservatively: a quantum efficiency of $60\%$, a reconciliation efficiency of $80\%$ and an excess noise of 0.005 or 0.01 units of shot noise. The superiority of the 8-dimensional protocol is quite clear: the secret key rate is higher by nearly an order of magnitude, and one can work with significantly larger modulation variances (the optimized variances are $V_A=0.3$ for the four-state protocol and $V_A=0.7$ for the 8-dimensional protocol).  

\begin{figure}[th]
  \centerline{
    \includegraphics[width=0.95\linewidth]{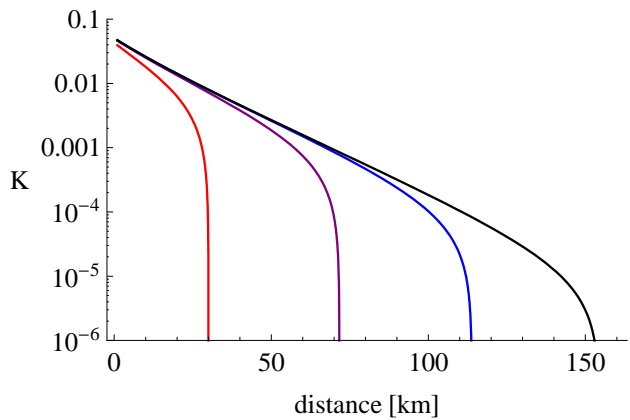}}
\caption{\label{KoctoFinite}(Color online) Non-asymptotic secret key rate $K$ for the 8-dimensional protocol, obtained for realistic  values:  excess noise $\xi = 0.005$, security parameter $\epsilon \approx 10^{-10}$, quantum efficiency of the detectors  $\eta = 60\%$, reconciliation efficiency $80\%$ for the BI-AWGN channel and transmission through a telecommunications fiber with 0.2 dB of loss per kilometer. Half the samples are used for parameter estimation. From left to right, the block length is equal to $10^{8}, 10^{10}, 10^{12}$ and $10^{14}$.}
\end{figure}

To confirm the robustness of the 8-dimensional protocol, we display in  Fig. \ref{KoctoFinite}  the secret key rate when finite-size effects are taken into account. The secret key rate is computed against collective attacks, as detailed in Ref. \cite{LGG10}.
Among various finite-size effects \cite{SR08}, the most crucial ones for continuous-variable protocols are clearly the imperfect reconciliation efficiency (which prevents the protocol with a Gaussian modulation to achieve key distribution over large distances) and the parameter estimation. While the reconciliation efficiency is taken care of by the 8-dimensional continuous modulation, the parameter estimation is quite sensitive for continuous-variable protocols. In fact, the real problem lies in the estimation of the excess noise $\xi$, which is very small compared to the shot noise, and thus hard to evaluate accurately.

In Fig. \ref{KoctoFinite}, all such finite-size effects are taken into account \cite{LGG10}.
The results are rather pessimistic, but remember that this is also true for all discrete-variable protocols \cite{CS09}, and the protocols presented here perform reasonably well in comparison. While exchanging $10^{14}$ quantum signals is rather unrealistic, exchanging $10^{10}$ or even $10^{11}$ signals is not completely out of reach of today's technology. Hence, our protocols allow the distribution of secret keys over distances of the order of 100 km,  taking into account all finite-size effects.

\section{Conclusion and perspectives}

In conclusion, we introduced continuous-variable QKD protocols with non-Gaussian modulations. We established the security of these protocols against arbitrary collective attacks, which implies their unconditional security in the asymptotic limit.

The four-state (the 8-dimensional) protocol appears optimal \emph{from a practical point of view} among all protocols using a homodyne (heterodyne) detection in the sense that it allows an efficient reconciliation while remaining as close as possible to the theoretically optimal Gaussian protocols \cite{GG02,WLB04}. 

The main open questions concern the status of the decoy states. First, it should be possible to prove the security of the protocols considered here without requiring any decoy states. This might come at the price of slightly degraded bounds (to take into account possible non Gaussian attacks). Second, without removing the decoy states, an important question is how well they should approximate the Gaussian distribution. Put otherwise, how indistinguishable should the distribution corresponding to key distillation be from the one used for parameter estimation? In particular, how should this distinguishability be taken into account in the overall security parameter of the protocol?

Finally, the most outstanding problem that remains for continuous-variable QKD protocol concerns general attacks. This question was partially answered with the derivation of a de Finetti-type theorem for quantum systems of infinite dimension \cite{RC09}. However, the bounds obtained there are not good enough to be used in practice. Hence, it seems crucial to see if the post-selection technique introduced in \cite{CKR09} can be adapted to continuous variables, since this technique is already known to provide much better (almost tight) bounds than the de Finetti theorem in the case of discrete variables \cite{SPS10}.

\section*{Acknowledgments}

The authors acknowledge fruitful discussions with Fr\'ed\'eric Grosshans, Norbert L\"utkenhaus and Renato Renner. This work was carried out in the framework of the ANR project SEQURE (ANR-07-SESU-011-01). AL received financial support from the EU ERC Starting grant PERCENT.


\begin{thebibliography}{32}
\expandafter\ifx\csname natexlab\endcsname\relax\def\natexlab#1{#1}\fi
\expandafter\ifx\csname bibnamefont\endcsname\relax
  \def\bibnamefont#1{#1}\fi
\expandafter\ifx\csname bibfnamefont\endcsname\relax
  \def\bibfnamefont#1{#1}\fi
\expandafter\ifx\csname citenamefont\endcsname\relax
  \def\citenamefont#1{#1}\fi
\expandafter\ifx\csname url\endcsname\relax
  \def\url#1{\texttt{#1}}\fi
\expandafter\ifx\csname urlprefix\endcsname\relax\def\urlprefix{URL }\fi
\providecommand{\bibinfo}[2]{#2}
\providecommand{\eprint}[2][]{\url{#2}}

\bibitem[{\citenamefont{Scarani et~al.}(2009)\citenamefont{Scarani,
  Bechmann-Pasquinucci, Cerf, Du{\v{s}}ek, L{\"u}tkenhaus, and Peev}}]{SBC08}
\bibinfo{author}{\bibfnamefont{V.}~\bibnamefont{Scarani}},
  \bibinfo{author}{\bibfnamefont{H.}~\bibnamefont{Bechmann-Pasquinucci}},
  \bibinfo{author}{\bibfnamefont{N.}~\bibnamefont{Cerf}},
  \bibinfo{author}{\bibfnamefont{M.}~\bibnamefont{Du{\v{s}}ek}},
  \bibinfo{author}{\bibfnamefont{N.}~\bibnamefont{L{\"u}tkenhaus}},
  \bibnamefont{and} \bibinfo{author}{\bibfnamefont{M.}~\bibnamefont{Peev}},
  \bibinfo{journal}{Rev. Mod. Phys.} \textbf{\bibinfo{volume}{81}},
  \bibinfo{eid}{1301} (\bibinfo{year}{2009}).

\bibitem[{\citenamefont{Cerf and Grangier}(2007)}]{CG07}
\bibinfo{author}{\bibfnamefont{N.~J.} \bibnamefont{Cerf}} \bibnamefont{and}
  \bibinfo{author}{\bibfnamefont{P.}~\bibnamefont{Grangier}},
  \bibinfo{journal}{Journal of the Optical Society of America B}
  \textbf{\bibinfo{volume}{24}}, \bibinfo{pages}{324} (\bibinfo{year}{2007}).

\bibitem[{\citenamefont{Grosshans and Grangier}(2002)}]{GG02}
\bibinfo{author}{\bibfnamefont{F.}~\bibnamefont{Grosshans}} \bibnamefont{and}
  \bibinfo{author}{\bibfnamefont{P.}~\bibnamefont{Grangier}},
  \bibinfo{journal}{Phys. Rev. Lett.} \textbf{\bibinfo{volume}{88}},
  \bibinfo{pages}{057902} (\bibinfo{year}{2002}).

\bibitem[{\citenamefont{Garc\'{\i}a-Patr\'{o}n and Cerf}(2006)}]{GC06}
\bibinfo{author}{\bibfnamefont{R.}~\bibnamefont{Garc\'{\i}a-Patr\'{o}n}}
  \bibnamefont{and} \bibinfo{author}{\bibfnamefont{N.~J.} \bibnamefont{Cerf}},
  \bibinfo{journal}{Phys. Rev. Lett.} \textbf{\bibinfo{volume}{97}},
  \bibinfo{pages}{190503} (\bibinfo{year}{2006}).

\bibitem[{\citenamefont{Navascu\'{e}s et~al.}(2006)\citenamefont{Navascu\'{e}s,
  Grosshans, and Ac\'{\i}n}}]{NGA06}
\bibinfo{author}{\bibfnamefont{M.}~\bibnamefont{Navascu\'{e}s}},
  \bibinfo{author}{\bibfnamefont{F.}~\bibnamefont{Grosshans}},
  \bibnamefont{and}
  \bibinfo{author}{\bibfnamefont{A.}~\bibnamefont{Ac\'{\i}n}},
  \bibinfo{journal}{Phys. Rev. Lett.} \textbf{\bibinfo{volume}{97}},
  \bibinfo{pages}{190502} (\bibinfo{year}{2006}).

\bibitem[{\citenamefont{Leverrier and Grangier}(2010{\natexlab{a}})}]{LG10b}
\bibinfo{author}{\bibfnamefont{A.}~\bibnamefont{Leverrier}} \bibnamefont{and}
  \bibinfo{author}{\bibfnamefont{P.}~\bibnamefont{Grangier}},
  \bibinfo{journal}{Phys. Rev. A} \textbf{\bibinfo{volume}{81}},
  \bibinfo{pages}{062314} (\bibinfo{year}{2010}{\natexlab{a}}).

\bibitem[{\citenamefont{Renner and Cirac}(2009)}]{RC09}
\bibinfo{author}{\bibfnamefont{R.}~\bibnamefont{Renner}} \bibnamefont{and}
  \bibinfo{author}{\bibfnamefont{J.~I.} \bibnamefont{Cirac}},
  \bibinfo{journal}{Phys. Rev. Lett.} \textbf{\bibinfo{volume}{102}},
  \bibinfo{pages}{110504} (\bibinfo{year}{2009}).

\bibitem[{\citenamefont{Lodewyck et~al.}(2007)\citenamefont{Lodewyck, Bloch,
  Garc\'{\i}a-Patr\'{o}n, Fossier, Karpov, Diamanti, Debuisschert, Cerf,
  Tualle-Brouri, McLaughlin et~al.}}]{LBG07}
\bibinfo{author}{\bibfnamefont{J.}~\bibnamefont{Lodewyck}},
  \bibinfo{author}{\bibfnamefont{M.}~\bibnamefont{Bloch}},
  \bibinfo{author}{\bibfnamefont{R.}~\bibnamefont{Garc\'{\i}a-Patr\'{o}n}},
  \bibinfo{author}{\bibfnamefont{S.}~\bibnamefont{Fossier}},
  \bibinfo{author}{\bibfnamefont{E.}~\bibnamefont{Karpov}},
  \bibinfo{author}{\bibfnamefont{E.}~\bibnamefont{Diamanti}},
  \bibinfo{author}{\bibfnamefont{T.}~\bibnamefont{Debuisschert}},
  \bibinfo{author}{\bibfnamefont{N.~J.} \bibnamefont{Cerf}},
  \bibinfo{author}{\bibfnamefont{R.}~\bibnamefont{Tualle-Brouri}},
  \bibinfo{author}{\bibfnamefont{S.~W.} \bibnamefont{McLaughlin}},
  \bibnamefont{et~al.}, \bibinfo{journal}{Phys. Rev. A}
  \textbf{\bibinfo{volume}{76}}, \bibinfo{eid}{042305} (\bibinfo{year}{2007}).

\bibitem[{\citenamefont{Fossier et~al.}(2009)\citenamefont{Fossier, Diamanti,
  Debuisschert, Villing, Tualle-Brouri, and Grangier}}]{FDD09}
\bibinfo{author}{\bibfnamefont{S.}~\bibnamefont{Fossier}},
  \bibinfo{author}{\bibfnamefont{E.}~\bibnamefont{Diamanti}},
  \bibinfo{author}{\bibfnamefont{T.}~\bibnamefont{Debuisschert}},
  \bibinfo{author}{\bibfnamefont{A.}~\bibnamefont{Villing}},
  \bibinfo{author}{\bibfnamefont{R.}~\bibnamefont{Tualle-Brouri}},
  \bibnamefont{and} \bibinfo{author}{\bibfnamefont{P.}~\bibnamefont{Grangier}},
  \bibinfo{journal}{New J. Phys.} \textbf{\bibinfo{volume}{11}}
  (\bibinfo{year}{2009}).

\bibitem[{\citenamefont{Leverrier et~al.}(2008)\citenamefont{Leverrier,
  All\'{e}aume, Boutros, Z\'{e}mor, and Grangier}}]{LAB08}
\bibinfo{author}{\bibfnamefont{A.}~\bibnamefont{Leverrier}},
  \bibinfo{author}{\bibfnamefont{R.}~\bibnamefont{All\'{e}aume}},
  \bibinfo{author}{\bibfnamefont{J.}~\bibnamefont{Boutros}},
  \bibinfo{author}{\bibfnamefont{G.}~\bibnamefont{Z\'{e}mor}},
  \bibnamefont{and} \bibinfo{author}{\bibfnamefont{P.}~\bibnamefont{Grangier}},
  \bibinfo{journal}{Phys. Rev. A} \textbf{\bibinfo{volume}{77}},
  \bibinfo{pages}{042325} (\bibinfo{year}{2008}).

\bibitem[{\citenamefont{Wolf et~al.}(2006)\citenamefont{Wolf, Giedke, and
  Cirac}}]{WGC06}
\bibinfo{author}{\bibfnamefont{M.~M.} \bibnamefont{Wolf}},
  \bibinfo{author}{\bibfnamefont{G.}~\bibnamefont{Giedke}}, \bibnamefont{and}
  \bibinfo{author}{\bibfnamefont{J.~I.} \bibnamefont{Cirac}},
  \bibinfo{journal}{Phys. Rev. Lett.} \textbf{\bibinfo{volume}{96}},
  \bibinfo{pages}{080502} (\bibinfo{year}{2006}).

\bibitem[{\citenamefont{Leverrier et~al.}(2009)\citenamefont{Leverrier, Karpov,
  Grangier, and Cerf}}]{LKG09}
\bibinfo{author}{\bibfnamefont{A.}~\bibnamefont{Leverrier}},
  \bibinfo{author}{\bibfnamefont{E.}~\bibnamefont{Karpov}},
  \bibinfo{author}{\bibfnamefont{P.}~\bibnamefont{Grangier}}, \bibnamefont{and}
  \bibinfo{author}{\bibfnamefont{N.~J.} \bibnamefont{Cerf}},
  \bibinfo{journal}{New J. Phys.} \textbf{\bibinfo{volume}{11}},
  \bibinfo{pages}{115009} (\bibinfo{year}{2009}).

\bibitem[{\citenamefont{Leverrier et~al.}(2010)\citenamefont{Leverrier,
  Grosshans, and Grangier}}]{LGG10}
\bibinfo{author}{\bibfnamefont{A.}~\bibnamefont{Leverrier}},
  \bibinfo{author}{\bibfnamefont{F.}~\bibnamefont{Grosshans}},
  \bibnamefont{and} \bibinfo{author}{\bibfnamefont{P.}~\bibnamefont{Grangier}},
  \bibinfo{journal}{Arxiv preprint 1005.0339}  (\bibinfo{year}{2010}).

\bibitem[{\citenamefont{Weedbrook et~al.}(2004)\citenamefont{Weedbrook, Lance,
  Bowen, Symul, Ralph, and Lam}}]{WLB04}
\bibinfo{author}{\bibfnamefont{C.}~\bibnamefont{Weedbrook}},
  \bibinfo{author}{\bibfnamefont{A.~M.} \bibnamefont{Lance}},
  \bibinfo{author}{\bibfnamefont{W.~P.} \bibnamefont{Bowen}},
  \bibinfo{author}{\bibfnamefont{T.}~\bibnamefont{Symul}},
  \bibinfo{author}{\bibfnamefont{T.~C.} \bibnamefont{Ralph}}, \bibnamefont{and}
  \bibinfo{author}{\bibfnamefont{P.~K.} \bibnamefont{Lam}},
  \bibinfo{journal}{Phys. Rev. Lett.} \textbf{\bibinfo{volume}{93}},
  \bibinfo{pages}{170504} (\bibinfo{year}{2004}).

\bibitem[{\citenamefont{Grosshans
  et~al.}(2003{\natexlab{a}})\citenamefont{Grosshans, Van~Assche, Wenger,
  Brouri, Cerf, and Grangier}}]{GVW03}
\bibinfo{author}{\bibfnamefont{F.}~\bibnamefont{Grosshans}},
  \bibinfo{author}{\bibfnamefont{G.}~\bibnamefont{Van~Assche}},
  \bibinfo{author}{\bibfnamefont{J.}~\bibnamefont{Wenger}},
  \bibinfo{author}{\bibfnamefont{R.}~\bibnamefont{Brouri}},
  \bibinfo{author}{\bibfnamefont{N.}~\bibnamefont{Cerf}}, \bibnamefont{and}
  \bibinfo{author}{\bibfnamefont{P.}~\bibnamefont{Grangier}},
  \bibinfo{journal}{Nature} \textbf{\bibinfo{volume}{421}},
  \bibinfo{pages}{238} (\bibinfo{year}{2003}{\natexlab{a}}).

\bibitem[{\citenamefont{Leverrier and Grangier}(2009)}]{LG09}
\bibinfo{author}{\bibfnamefont{A.}~\bibnamefont{Leverrier}} \bibnamefont{and}
  \bibinfo{author}{\bibfnamefont{P.}~\bibnamefont{Grangier}},
  \bibinfo{journal}{Phys. Rev. Lett.} \textbf{\bibinfo{volume}{102}},
  \bibinfo{pages}{180504} (\bibinfo{year}{2009}).

\bibitem[{\citenamefont{Zhao et~al.}(2009)\citenamefont{Zhao, Heid, Rigas, and
  L\"{u}tkenhaus}}]{ZHR09}
\bibinfo{author}{\bibfnamefont{Y.-B.} \bibnamefont{Zhao}},
  \bibinfo{author}{\bibfnamefont{M.}~\bibnamefont{Heid}},
  \bibinfo{author}{\bibfnamefont{J.}~\bibnamefont{Rigas}}, \bibnamefont{and}
  \bibinfo{author}{\bibfnamefont{N.}~\bibnamefont{L\"{u}tkenhaus}},
  \bibinfo{journal}{Phys. Rev. A} \textbf{\bibinfo{volume}{79}},
  \bibinfo{pages}{012307} (\bibinfo{year}{2009}).

\bibitem[{\citenamefont{Leverrier and Grangier}(2010{\natexlab{b}})}]{LG10}
\bibinfo{author}{\bibfnamefont{A.}~\bibnamefont{Leverrier}} \bibnamefont{and}
  \bibinfo{author}{\bibfnamefont{P.}~\bibnamefont{Grangier}},
  \bibinfo{journal}{Arxiv preprint 1002.4083}
  (\bibinfo{year}{2010}{\natexlab{b}}).

\bibitem[{\citenamefont{Leverrier and Grangier}(2010{\natexlab{c}})}]{LG10c}
\bibinfo{author}{\bibfnamefont{A.}~\bibnamefont{Leverrier}} \bibnamefont{and}
  \bibinfo{author}{\bibfnamefont{P.}~\bibnamefont{Grangier}},
  \bibinfo{journal}{Arxiv preprint 1005.0328}
  (\bibinfo{year}{2010}{\natexlab{c}}).

\bibitem[{\citenamefont{Devetak and Winter}(2005)}]{DW05}
\bibinfo{author}{\bibfnamefont{I.}~\bibnamefont{Devetak}} \bibnamefont{and}
  \bibinfo{author}{\bibfnamefont{A.}~\bibnamefont{Winter}}, in
  \emph{\bibinfo{booktitle}{Proc. R. Soc. A}} (\bibinfo{year}{2005}), vol.
  \bibinfo{volume}{461}, pp. \bibinfo{pages}{207--235}.

\bibitem[{\citenamefont{Grosshans
  et~al.}(2003{\natexlab{b}})\citenamefont{Grosshans, Cerf, Wenger,
  Tualle-Brouri, and Grangier}}]{GCW03}
\bibinfo{author}{\bibfnamefont{F.}~\bibnamefont{Grosshans}},
  \bibinfo{author}{\bibfnamefont{N.}~\bibnamefont{Cerf}},
  \bibinfo{author}{\bibfnamefont{J.}~\bibnamefont{Wenger}},
  \bibinfo{author}{\bibfnamefont{R.}~\bibnamefont{Tualle-Brouri}},
  \bibnamefont{and} \bibinfo{author}{\bibfnamefont{P.}~\bibnamefont{Grangier}},
  \bibinfo{journal}{{Quantum Information and Computation}}
  \textbf{\bibinfo{volume}{3}}, \bibinfo{pages}{535}
  (\bibinfo{year}{2003}{\natexlab{b}}).

\bibitem[{\citenamefont{Scarani and Renner}(2008)}]{SR08}
\bibinfo{author}{\bibfnamefont{V.}~\bibnamefont{Scarani}} \bibnamefont{and}
  \bibinfo{author}{\bibfnamefont{R.}~\bibnamefont{Renner}},
  \bibinfo{journal}{Phys. Rev. Lett.} \textbf{\bibinfo{volume}{100}},
  \bibinfo{pages}{200501} (\bibinfo{year}{2008}).

\bibitem[{\citenamefont{Cai and Scarani}(2009)}]{CS09}
\bibinfo{author}{\bibfnamefont{R.~Y.~Q.} \bibnamefont{Cai}} \bibnamefont{and}
  \bibinfo{author}{\bibfnamefont{V.}~\bibnamefont{Scarani}},
  \bibinfo{journal}{New J. Phys.} \textbf{\bibinfo{volume}{11}}
  (\bibinfo{year}{2009}).

\bibitem[{\citenamefont{Christandl et~al.}(2009)\citenamefont{Christandl,
  K\"{o}nig, and Renner}}]{CKR09}
\bibinfo{author}{\bibfnamefont{M.}~\bibnamefont{Christandl}},
  \bibinfo{author}{\bibfnamefont{R.}~\bibnamefont{K\"{o}nig}},
  \bibnamefont{and} \bibinfo{author}{\bibfnamefont{R.}~\bibnamefont{Renner}},
  \bibinfo{journal}{Phys. Rev. Lett.} \textbf{\bibinfo{volume}{102}},
  \bibinfo{eid}{020504} (\bibinfo{year}{2009}).

\bibitem[{\citenamefont{Sheridan et~al.}(2010)\citenamefont{Sheridan, Le, and
  Scarani}}]{SPS10}
\bibinfo{author}{\bibfnamefont{L.}~\bibnamefont{Sheridan}},
  \bibinfo{author}{\bibfnamefont{T.~P.} \bibnamefont{Le}}, \bibnamefont{and}
  \bibinfo{author}{\bibfnamefont{V.}~\bibnamefont{Scarani}},
  \bibinfo{journal}{New Journal of Physics} \textbf{\bibinfo{volume}{12}},
  \bibinfo{pages}{123019} (\bibinfo{year}{2010}).

\bibitem[{\citenamefont{Grangier et~al.}(1998)\citenamefont{Grangier, Levenson,
  and Poizat}}]{GLP98}
\bibinfo{author}{\bibfnamefont{P.}~\bibnamefont{Grangier}},
  \bibinfo{author}{\bibfnamefont{J.}~\bibnamefont{Levenson}}, \bibnamefont{and}
  \bibinfo{author}{\bibfnamefont{J.}~\bibnamefont{Poizat}},
  \bibinfo{journal}{Nature} \textbf{\bibinfo{volume}{396}},
  \bibinfo{pages}{537} (\bibinfo{year}{1998}), ISSN \bibinfo{issn}{0028-0836}.

\bibitem[{\citenamefont{Wyner}(1975)}]{wyn75}
\bibinfo{author}{\bibfnamefont{A.}~\bibnamefont{Wyner}}, \bibinfo{journal}{Bell
  System Technical Journal} \textbf{\bibinfo{volume}{54}},
  \bibinfo{pages}{1355} (\bibinfo{year}{1975}).

\bibitem[{\citenamefont{Richardson et~al.}(2001)\citenamefont{Richardson,
  Shokrollahi, and Urbanke}}]{RSU01}
\bibinfo{author}{\bibfnamefont{T.}~\bibnamefont{Richardson}},
  \bibinfo{author}{\bibfnamefont{M.}~\bibnamefont{Shokrollahi}},
  \bibnamefont{and} \bibinfo{author}{\bibfnamefont{R.}~\bibnamefont{Urbanke}},
  \bibinfo{journal}{IEEE Transactions on Information Theory}
  \textbf{\bibinfo{volume}{47}}, \bibinfo{pages}{619} (\bibinfo{year}{2001}).

\bibitem[{\citenamefont{Richardson and Urbanke}(2002)}]{RU02}
\bibinfo{author}{\bibfnamefont{T.}~\bibnamefont{Richardson}} \bibnamefont{and}
  \bibinfo{author}{\bibfnamefont{R.}~\bibnamefont{Urbanke}},
  \bibinfo{journal}{Workshop honoring Prof. Bob McEliece on his 60th birthday}
  pp. \bibinfo{pages}{24--25} (\bibinfo{year}{2002}).

\bibitem[{\citenamefont{Leverrier}(2009)}]{lev09}
\bibinfo{author}{\bibfnamefont{A.}~\bibnamefont{Leverrier}}, Ph.D. thesis,
  \bibinfo{school}{{Ecole Nationale Sup\'erieure des T\'el\'ecommunications}}
  (\bibinfo{year}{2009}),
  \urlprefix\url{http://tel.archives-ouvertes.fr/tel-00451021}.

\bibitem[{\citenamefont{Leverrier and Cerf}(2009)}]{LC09}
\bibinfo{author}{\bibfnamefont{A.}~\bibnamefont{Leverrier}} \bibnamefont{and}
  \bibinfo{author}{\bibfnamefont{N.~J.} \bibnamefont{Cerf}},
  \bibinfo{journal}{Phys. Rev. A} \textbf{\bibinfo{volume}{80}},
  \bibinfo{eid}{010102} (\bibinfo{year}{2009}).

\bibitem[{\citenamefont{Mezzadri}(2006)}]{mez06}
\bibinfo{author}{\bibfnamefont{F.}~\bibnamefont{Mezzadri}},
  \bibinfo{journal}{Arxiv preprint math-ph/0609050}  (\bibinfo{year}{2006}).

\end{thebibliography}

\appendix

\section{Linear quantum channels}
\label{linear}

We shall define a linear quantum channel by the input-output  relations of the quadrature operators in Heisenberg representation : 
\begin{eqnarray}
X_{out} = g_X X_{in} + B_X  \nonumber \\
P_{out} = g_P  P_{in} + B_P
\end{eqnarray}
where the added noises  $B_X$, $B_P$ are uncorrelated with the input quadratures $X_{in}$, $P_{in} $. 
Such relations have been extensively used for instance in the context of Quantum Non-Demolition (QND) measurements of continuous variables \cite{GLP98}, and they are closely related to the linearized approximation commonly used in quantum optics. Gaussian channels (channels that preserve the Gaussianity of the states) are usual examples of linear quantum channels. However, linear quantum channels may also be non-Gaussian, this will be the case for instance if the added noises $B_X$, $B_P$ are non-Gaussian. 

For our purpose, the main advantage of a linear quantum channel is that it will be characterized by transmission coefficients $T_X = g_X^2$, $T_P = g_P^2$, and by the variances of the added noises $B_X$ and $B_P$. These quantities can be determined even if the modulation used by Alice is non-Gaussian, with the same measured values as when the modulation is Gaussian (because these values are intrinsic properties of the channel). The relevant covariance matrix can then be easily determined, and Eve's information can be bounded by using the Gaussian optimality theorem. This justifies the approach taken in refs. \cite{LG09,LG10}, but unfortunately this is not fully general,  contrary to the proof of the present paper.

\section{Efficient reverse reconciliation}
\label{reduction}

The goal of this section is to explain how an efficient reconciliation can be achieved for the various modulation schemes considered in this article, for arbitrarily low SNR.

Reconciliation in a QKD protocol is very similar to the problem of channel coding (that is, transmitting information efficiently and reliably on a noisy communication channel) with the additional constraint that the input of the channel is not chosen by Alice but instead randomly picked from a given probability distribution corresponding to the modulation scheme. In particular, the usual task is the following: Alice and Bob are given two $n$-dimensional real vectors ${\bf x}$ and ${\bf y}$ and their goal is to agree on a common bit string ${\bf u}$. A supplementary constraint when dealing with reverse reconciliation is that ${\bf u}$ should be a function (possibly randomized) of ${\bf y}$, and that all public communication should be from Bob to Alice.

There exists a standard technique for reducing the problem of reconciliation to the one of channel coding, namely coset coding introduced by Wyner \cite{wyn75}. The idea is that Bob will use an additional public (but authenticated) channel to describe a function $f$ such $f({\bf y}) = {\bf u}$. Alice can then apply this function to her vector and obtain ${\bf v} := f({\bf x})$. In the case of coset coding, the description of $f$ is simply a translation corresponding to the syndrome of $y$ for a linear error correcting code $C$. This gives rise to a virtual communication channel with input ${\bf u}$ and output ${\bf v}$ for which one can apply standard channel coding techniques. 

One case where reconciliation can be performed very efficiently occurs when the virtual channel is a binary-input additive white Gaussian noise (BI-AWGN) channel meaning that the coordinates of ${\bf u}$ and ${\bf v}$ are related through
\begin{equation}
v_i = u_i + w_i,
\end{equation}
where $u_i \in \{-1,1\}$ and $w_i$ is a centered normal random variable.

For experimental realizations of continuous-variable QKD, the quantum channel always behaves in very good approximation like a Gaussian channel and the BI-AWGN channel is therefore the model of interest here. In this case, it is possible to show that the existence of an efficiently computable function $f$ is possible only for very specific modulation schemes, namely the cases where $d$-uplets of $x_i$ are distributed uniformly on the unit sphere in dimension 1, 2, 4 or 8 \cite{LAB08}. The case $d=1$ corresponds to a binary modulation, that is, the four-state protocol (which indeed displays a binary modulation for \emph{each} quadrature) ; the case $d=8$ corresponds to the 8-dimensional modulation. 

Let us therefore consider $d$-uplets ${\bf x^d}$ and ${\bf y^d}$, with ${\bf x^d }\sim \mathcal{U}(\mathcal{S}^{d-1})$ and ${\bf y^d} =  {\bf x^d} + {\bf z^d}$ with ${\bf z^d} \sim \mathcal{N}(0, \sigma^2)^d$. Without loss of generality, we restrict our attention to the case where the transmission is 1.

For dimensions 1, 2, 4 and 8, the unit sphere $\mathcal{S}^{d-1}$ has a division algebra. In particular, a $d$-dimensional vector can be identified with an element of $\mathbbm{R}^d$, that is, a real number ($d=1$), a complex number ($d=2$), a quaternion ($d=4$) or an octonion ($d=8$). Therefore, both multiplication and division are well defined in this context.

Bob chooses a random element ${\bf u^d} \in \{-1/\sqrt{d},1/\sqrt{d}\}^d$ with the uniform distribution on the $d$-dimensional hypercube and sends the variable ${\bf t^d} := {\bf u^d} {\bf y^d} $ to Alice (through the classical channel).
Alice computes ${\bf v^d} :=  {\bf t^d} \left({\bf x^d}\right)^{-1}$ which is possible because $\mathcal{S}^{d-1}$ is a division algebra.
We now wish to prove that the channel ${\bf u^d} \rightarrow {\bf v^d}= {\bf u^d} + {\bf w^d}$ is a BI-AWGN channel.
Let us characterize the noise ${\bf w^d}$ on this virtual channel:
\begin{eqnarray}
{\bf w^d} &\equiv & {\bf v^d} - {\bf u^d} \\
&=& {\bf t^d} \left({\bf x^d}\right)^{-1} - {\bf u^d} \\
&=& {\bf u^d} {\bf y^d} \left({\bf x^d}\right)^{-1} - {\bf u^d} \\
&=& {\bf u^d} ({\bf y^d} \left({\bf x^d}\right)^{-1}-1) \\
&=& {\bf u^d} (({\bf x^d} +{\bf z^d}) \left({\bf x^d}\right)^{-1}-1)\\
&=& {\bf u^d} {\bf z^d} \left({\bf x^d}\right)^{-1}.
\end{eqnarray}
Since ${\bf u^d}$ and $\left({\bf x^d}\right)^{-1}$ are simply rotations on $\mathcal{S}^{d-1}$, one concludes that ${\bf w^d} \sim \mathcal{N}(0,\sigma^2)^{\otimes d}$, which proves that the virtual channel  ${\bf u^d} \rightarrow {\bf v^d}$ is indeed a BI-AWGN channel, for which efficient error correcting codes are available.

If $n= d \times m$, Alice and Bob simply divide their data into $m$ $d$-uplets and proceed as described above. All is left to do is to use coset coding to finish the reconciliation: Bob sends the syndrome of ${\bf u^n}$ for a linear code Alice and he agreed on beforehand. Alice simply decodes her word ${\bf v^n}$ in the coset code defined by the syndrome. This can be done very efficiently with capacity approaching codes such as low-density parity-check (LDPC) codes \cite{RSU01}. If the SNR is very low, then one can work with a concatenation of a low rate LDPC code (such as a multi-edge LDPC code for instance \cite{RU02}) with a repetition code. This simple technique allows one to obtain good error correcting codes of arbitrarily low rate (see Chapter 5.2 of Ref. \cite{lev09} for more details concerning this concatenation technique).

\section{Decoys with coherent states} 
\label{decoys_coherent}

In the entanglement-based version of the protocol, one has to apply the generalized measurements $\{\Pi_d, \mathbbm{1}-\Pi_d\}$. In the prepare and measure scenario, it is therefore necessary for Alice to send states which are compatible with these measurements. The states corresponding to the operator $\Pi_d$ are not a problem, since by construction, they correspond to the modulation used to distill the key, that is, coherent states drawn uniformly on the sphere in $d$ dimensions.
The states corresponding to the operator $\mathbbm{1}-\Pi_d$ might be a little bit more problematic in the sense that they are not usually easy to produce experimentally. Ideally, one would like to be able to produce these states simply by modulating coherent states (which are the only states simple enough to allow for a realistic QKD protocol).
In particular, if Alice applies the measurements $\{\Pi_d, \mathbbm{1}-\Pi_d\}$, then one obtains the following relation:
\begin{equation}
\label{def_decoy}
\sigma_{\mathrm{G}}^d=  p\, \sigma_{\mathrm{key}}^d + (1-p) \, \sigma_{\mathrm{decoy}}^d,
\end{equation}
with $p = p_d^{\mathrm{succ}}$.
However, in this case, the state $\sigma_{\mathrm{decoy}}^d$ does not have a positive $P$-function, meaning that it cannot be obtained as a mixture of coherent states.

We now present two different techniques to deal with this problem: either one replaces $\sigma_{\mathrm{key}}^d$ with a noisy version (see Appendix \ref{noisy}) or one relaxes Equation \ref{def_decoy} and considers instead an approximate version of the decoy states (see Appendix \ref{optimization}).

\subsection{Perfect decoys with noisy signal}
\label{noisy}

\begin{figure}[!ht]
\centerline{
\includegraphics[width=0.95\linewidth]{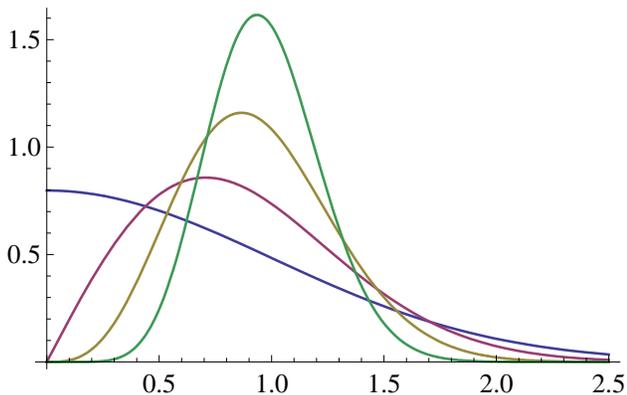} 
}
\caption{\label{chi_distrib}(Color online) Probability density functions for the radius of ${\bf \alpha}$ in the $d$-dimensional protocol for a Gaussian modulation. From least peaked to most peaked, $d=1,2,4$ and $8$.}
\end{figure}

Let us consider the prepare and measure version of the $d$-dimensional protocol. In this case, the Gaussian modulation can be seen as sending $d/2$ coherent states $|\alpha_1 + i \alpha_2\rangle, \cdots, |\alpha_{d-1}+  i \alpha_{d}\rangle$ such that the random variables $\alpha_i$ are independent and identically distributed centered normal variables. Without loss of generality, we consider variables with variance $1/d$. Taking advantage of the rotational invariance of the Gaussian distribution, one can equivalently choose the random vector ${\bf \alpha} := (\alpha_1, \cdots, \alpha_d )$ by first picking uniformly a random point of the unit sphere $\mathcal{S}^{d-1}$ in $d$ dimensions, and drawing the radius $r:= \sqrt{\sum \alpha_i^2}$ of the vector ${\bf \alpha}$ from a chi distribution with $d$ degrees of freedom. In particular, the probability density function of $r$ is:
\begin{equation}
f(r,d)= \frac{2 \left(d/2\right)^{d/2} r^{d-1}e^{-d r^2/2}}{\left(d/2-1 \right)!}.	
\end{equation} 
The probability distributions corresponding to $d=1,2,4$ and $8$ are displayed on Figure \ref{chi_distrib}. In particular, it should be noted that they become more and more peaked as the dimension $d$ increases. In this picture, the probability distribution corresponding to the key, that is $\sigma_\mathrm{key}^d$, is by construction a Dirac distribution centered in 1.

The first approach we investigate aims at satisfying Equation \ref{def_decoy} exactly while allowing for a positive (and non-negligible) probability $p$ of sending a signal state. This is done by considering slightly noisy versions of the true signal modulation.

In particular, one chooses two parameters $\gamma_\mathrm{min}\in [0,1]$ and $\gamma_\mathrm{max} \geq 1$ and defines the states used for key distillation as the ones with a radius bounded by these two parameters: $\gamma_\mathrm{min} \leq r \leq \gamma_\mathrm{max}$. The decoy states then simply correspond to the remaining states. Provided that $\gamma_\mathrm{min}$ and $\gamma_\mathrm{max}$ are close enough to 1, the penalty imposed by this noise, compared to the ideal case where the key modulation is strictly equal to 1, is negligible in terms of reconciliation efficiency. 

On the other hand, one should not choose values too close to 1, otherwise the probability $p$ that a given state can be used for key distillation will be very small:
\begin{equation}
p = \int_{\gamma_\mathrm{min}}^{\gamma_\mathrm{max}} \, f(r,d) \, \mathrm{d} r.
\end{equation}

Hence, optimizing the values of $\gamma_\mathrm{min}$ and $\gamma_\mathrm{max}$ should be seen as a tradeoff between the penalty imposed on reconciliation efficiency and the probability that a given state can be used for key distillation.

Note also that with this approach, the state $\sigma_\mathrm{key}^d$ is not longer described as in Appendix \ref{covariance}. In particular, the covariance matrix of the new state is a little bit different from the one presented in Appendix \ref{covariance}. We do not give an explicit derivation of the new covariance matrix here, but we point out that because the new state used for the key is actually closer to a Gaussian modulation, the Holevo information between Bob and Eve can still be safely bounded using the covariance matrix given in Eq. \ref{Gamma_key}.

We now give a second approach to the problem of approximating decoy states with coherent states.

\subsection{Approximate decoys with noiseless signal}
\label{optimization}

Our goal is still to achieve the following equality:
\begin{equation}
\label{def_decoy2}
\sigma_{\mathrm{G}}^d= p \, \sigma_{\mathrm{key}}^d + (1-p) \, \sigma_{\mathrm{decoy}}^d, 
\end{equation}
but this time, without considering a noisy version of $\sigma_{\mathrm{key}}^d $.
If one chooses $p = p_d^{\mathrm{succ}}$ as defined in Appendix \ref{measurement}, then the state $\sigma_{\mathrm{decoy}}^d$ does not have a positive $P$-function, meaning that it cannot be obtained as a mixture of coherent states. 
Fortunately, $P$-functions can be regularized rather well, and for our purpose, it is sufficient to find a state $\sigma_{\mathrm{decoy}}^d$ with a non negative $P$-function such that Equation \ref{def_decoy2} only holds \emph{approximately}. Here, approximately should be understood in terms of the trace distance.

More precisely, we are interested in finding a probability distribution (hence non negative) $P(\alpha)$ such that
\begin{equation}
\label{approx}
||\sigma_{\mathrm{G}}^d- p \, \sigma_{\mathrm{key}}^d - (1-p) \, 
 \int P(\alpha) |\alpha\rangle \! \langle \alpha|\,  \mathrm{d} \alpha||_1 \leq \epsilon
\end{equation}
for a value of $\epsilon$ sufficiently small. 

If Eq. \ref{approx} holds and Alice uses the modulation $P(\alpha)$ for the decoy states, then the probability that Eve can distinguish the states used for key distillation from the ones used for parameter estimation is upper bounded by $\epsilon$.

Here, we do not give a solution for the problem of finding the best distribution $P$ compatible with a success probability $p$ and the error $\epsilon$, but we point out that the usual optimization tools (for instance the optimization toolbox of Matlab) allow one to find very good instances of $P$. For example, for the 2-dimensional protocol and $\alpha=0.5$, we could obtain a value of $\epsilon$ less than $10^{-5}$ for $p=1/2$ with a distribution $P$ corresponding to a sum of 6 Dirac distributions, that is a mixture of 6 particular coherent states. 

The natural question that arises here is how good should the approximation be? Is a trace distance equal to $10^{-5}$ sufficient to guarantee a reasonable level of security? Or should one aim for a value of $10^{-10}$?

Note that if the approximation is not perfect, it means that Eve might have a (very) small probability to distinguish between the states used for the key distillation and those used for parameter estimation. However, discriminating between these two modulations does not appear to be a good solution for Eve as this would induce a lot of phase noise in the signal: indeed, because all the modulations considered here are phase-invariant, the optimal discrimination procedure consists in projecting the states onto Fock states, thereby erasing all the phase information.
For this reason, a trace distance of $10^{-5}$ between $\rho_{\mathrm{decoy}}$ and its approximation is very likely to be sufficient for any practical implementation.

\section{Measurement operator}
\label{measurement}

We now describe the general measurement $\{\Pi_d, \mathbbm{1}-\Pi_d\}$ performed by Alice to prepare the state $\rho_\mathrm{key}^0$ from a two-mode squeezed vacuum. The state to which this measurement is applied is a (possibly multimodal) two-mode squeezed vacuum as described in Table \ref{table_resource}.

\begin{table}
\begin{center}
\begin{tabular}{|c|c|c|}
\hline
$d$  & Protocol & Resource state   \\
\hline \hline
 1 & 4-state & $|\mathrm{EPR} \rangle $ \\
\hline
 2 & 2-dim & $|\mathrm{EPR} \rangle $ \\
\hline
 4 & 4-dim & $|\mathrm{EPR} \rangle^{\otimes 2} $ \\
\hline
 8 & 8-dim & $|\mathrm{EPR} \rangle^{\otimes 4} $ \\
\hline
\end{tabular}
\label{table_resource}
\caption{Parameterization of the various protocols. Parameter $d$ corresponds to the number of quadratures that should be processed together.}
\end{center}
\end{table}

\subsubsection{Four-state protocol: $d=1$}

The operator $\Pi_1$ is defined as $\Pi_1 = M_1^{\dagger}M_1$ with 
\begin{equation}
M_1 = m_1 \sum_{k=0}^3 |\psi_k\rangle\!\langle e_k|
\end{equation}
and
\begin{equation}
m_1:= \frac{e^{(1+\alpha^2)/2}}{2} \sqrt{\frac{\lfloor 1+\alpha^2 \rfloor !}{(1+\alpha^2)^{\lfloor 1+\alpha^2 \rfloor}}}.
\end{equation}
The states $|\psi_k\rangle$ are defined in Appendix \ref{covariance} and
\begin{equation}
|e_k\rangle=e^{-\beta^2/2}\sum_{n=0}^\infty \frac{\beta_k^{* n}}{\sqrt{n!}}|n\rangle
\end{equation}
with $\beta = \sqrt{1+\alpha^2}$ and $\mathrm{Arg}(\beta_k)=\mathrm{Arg}(\alpha_k)$.

When performing the general measurement $\{ \Pi_1, \mathbbm{1}-\Pi_1\}$, conditioned on the result corresponding to $\Pi_1$, the state $\rho$ is transformed into $\rho'$:
\begin{equation}
\rho \longrightarrow\rho' := \frac{M_1 \rho M^{\dagger}_1}{\mathrm{tr} M_1 \rho M^{\dagger}_1}.
\end{equation}

Let us consider the state $\rho_{\mathrm{G}}^0:= |\mathrm{EPR}\rangle\!\langle \mathrm{EPR}|$ with
\begin{equation}
|\mathrm{EPR}\rangle = \sum_{n=0}^{\infty} \sqrt{\frac{\alpha^{2n}}{(1+\alpha^2)^{n+1}}} |n\rangle |n\rangle.
\end{equation}
Conditioned on the result $\Pi_1$, one obtains
\begin{equation}
M_1 \rho_{\mathrm{G}} M_1^{\dagger} = \frac{4m_1^2 e^{-(\alpha^2+1)}}{\alpha^2+1}  |\Psi_1\rangle \!\langle \Psi_1|
\end{equation}
with
\begin{equation}
|\Psi_1\rangle = \frac{1}{2} \sum_{k}^{3}  |\psi_k\rangle  |\alpha_k \rangle.
\end{equation}

The condition $\Pi \leq \mathbbm{1}$ leads to $m_1 \leq m_1^{\mathrm{max}}$ with 
\begin{equation}
m_1^{\mathrm{max}} := \frac{e^{(1+\alpha^2)/2}}{2} \sqrt{\frac{\lfloor 1+\alpha^2 \rfloor !}{(1+\alpha^2)^{\lfloor 1+\alpha^2 \rfloor}}}.
\end{equation}

\begin{figure}[!ht]
\centerline{
\includegraphics[width=0.95\linewidth]{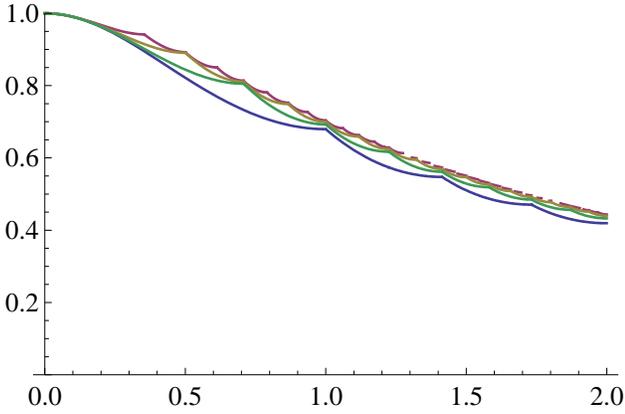} 
}
\caption{\label{proba_Pi}(Color online) Probability of success of the measurement $\Pi_d$ as a function of $\alpha$. From bottom to top, $d=1,2,4$ and $8$.}
\end{figure}

The probability of obtaining the result corresponding to $\Pi_1$, meaning successfully creating a state $|\Psi_1\rangle$ is 
\begin{equation}
p_1^{\mathrm{succ}}= \mathrm{tr} M_1 \rho M_1^{\dagger} =\frac{\lfloor 1+\alpha^2 \rfloor !}{(1+\alpha^2)^{\lfloor 2+\alpha^2 \rfloor}} 
\end{equation}
and is displayed on Figure \ref{proba_Pi} as a function of $\alpha$.

\subsubsection{Continuous modulations: $d=2,4,8$}

For $d=2,4,8$, one has:
\begin{equation}
|\Psi_d\rangle = \sum_{k=0}^\infty \sqrt{f_n(k)} |\psi_k^d\rangle
\end{equation}
where
\begin{equation}
f_n(k)= e^{-n \alpha^2} \frac{n^k \alpha^{2k}}{k!}
\end{equation}
and 
\begin{eqnarray*}
|\psi_k^2\rangle &:=& |k\rangle | k \rangle\\
|\psi_k^4\rangle &:=& \frac{1}{\sqrt{k+1}} \sum_{k_1 =0}^k |k_1,k-k_1\rangle | k_1,k-k_1 \rangle\\
|\psi_k^8\rangle &:=& \frac{1}{\sqrt{{k+3 \choose 3}}} \sum_{\sum_i k_i = k } |k_1,k_2,k_3,k_4\rangle | k_1,k_2,k_3,k_4 \rangle
\end{eqnarray*}

One also has:
\begin{eqnarray*}
|\mathrm{EPR} \rangle &=& \sum_{k=0}^\infty \sqrt{g_2(k)} |\psi_k^2\rangle \\
|\mathrm{EPR}\rangle^{\otimes 2} &=& \sum_{k=0}^\infty \sqrt{g_4(k)} |\psi_k^4\rangle \\
|\mathrm{EPR} \rangle^{\otimes 4} &=&  \sum_{k=0}^\infty \sqrt{g_8(k)} |\psi_k^8\rangle 
\end{eqnarray*}
with
\begin{equation}
g_n(k)= {n+k-1 \choose k} \frac{\alpha^{2k}}{(1+\alpha^2)^{n+k}}.
\end{equation}
$g_n$ is a negative binomial distribution $\mathrm{NB}\left(d,\frac{\alpha^2}{1+\alpha^2}\right)$.
Let us define the operators $\Pi_2, \Pi_4$ and $\Pi_8$ as
\begin{equation}
\Pi_d = \pi_d \sum_{k=0}^{\infty} \frac{f_d(k)}{g_d(k)} \, \mathrm{tr}_B |\psi_k^d\rangle \! \langle \psi_k^d|
\end{equation}
where $\pi_d$ is given by:
\begin{equation}
\pi_d(\alpha) = \min_{k \in \mathbb{N}} \frac{g(k)}{f(k)}
\end{equation}
ensuring that $\Pi_d$ is a genuine POVM element.
It is straightforward to check that the probability of success of the measurement is 
\begin{equation}
p_d^{\mathrm{succ}} = \frac{g(\lceil \alpha^2 d \rceil)}{f(\lceil 
\alpha^2 d \rceil)}.
\end{equation}

\section{Covariance matrices}
\label{covariance}

In this appendix, we derive the covariance matrices of the states corresponding to the four-state protocol, which is optimal with a homodyne detection, and the 8-dimensional protocol which is optimal for a heterodyne detection. The covariance matrices corresponding to the other (suboptimal) choices of modulation can be found with a similar technique.

Let us note $|\Psi_d\rangle$ for $d \in \{1,2,4,8\}$ the state used for the key distillation in each protocol, \emph{i.e.} $\rho_\mathrm{key}^0 = |\Psi_d\rangle \!\langle \Psi_d|$ and
$|\Psi_1\rangle$ is the initial bipartite state for the four-state protocol whereas $|\Psi_8\rangle$ corresponds to the 8-dimensional protocol. 
The two-mode squeezed vacuum is noted as $|\mathrm{EPR}\rangle$.

\label{covariance_matrix}

\subsection{Four-state protocol: $d=1$}
\label{four-state}

Let us use the notation $|\alpha_k\rangle := |\alpha e^{(2k+1)i \pi/4}\rangle$ for $k \in \{0,1,2,3\}$ and $\alpha >0$.

In the four-state protocol, the state $\rho_\mathrm{key}^0$ is a pure state $|\Psi_1\rangle$ defined as
\begin{eqnarray}
|\Psi_1\rangle &=& \sum_{k=0}^3 \sqrt{\lambda_k} |\phi_k\rangle |\phi_k \rangle \\
&=& \frac{1}{2}\sum_{k=0}^3|\psi_k\rangle |\alpha_k\rangle
\end{eqnarray}
where  
\begin{eqnarray}
|\phi_k\rangle &=& \frac{e^{-\alpha^2/2}}{\sqrt{\lambda_k}} \sum_{n=0}^{\infty} (-1)^n \frac{\alpha^{4n+k}}{\sqrt{(4n+k)!}} |4n+k\rangle\\
|\psi_k\rangle &=& \frac{1}{2} \sum_{m=0}^3 e^{i(1+2k)m \pi/4 }|\phi_m\rangle
\end{eqnarray}
for $k \in \{0,1,2,3\}$ and
\begin{equation}
\left\{
\begin{array}{lll}
\lambda_{0,2} &= & \frac{1}{2} e^{-\alpha^2}\left( \cosh(\alpha^2) \pm \cos(\alpha^2) \right) \\
\lambda_{1,3} &= & \frac{1}{2} e^{-\alpha^2}\left( \sinh(\alpha^2) \pm \sin(\alpha^2) \right) 
\end{array}
\right.
\end{equation}

Let us note $a$ and $b$ the annihilation operators on the two modes.
Applying $a$ to $|\phi_k\rangle$ gives:
\begin{equation}
a |\phi_k\rangle = \alpha \frac{\sqrt{\lambda_{k-1}}}{\sqrt{\lambda_k}} |\phi_{k-1}\rangle
\end{equation}
for $k \in \{1,2,3\}$ and
\begin{equation}
a |\phi_0\rangle = -\alpha \frac{\sqrt{\lambda_{3}}}{\sqrt{\lambda_0}} |\phi_{3}\rangle.
\end{equation}

Let us compute the covariance matrix $\Gamma_1$ of the bipartite state $|\Psi_1\rangle$. It has the following form:
\begin{equation}  
 \Gamma_1 = 
  \left(
\begin{array}{cc}
    X \mathbbm{1}_2  & Z_1 \, \sigma_z\\
    Z_1\,  \sigma_z & Y \mathbbm{1}_2\\
 \end{array}
\right)
\end{equation}

One has:
\begin{eqnarray}
X=Y&=&\langle \Psi_1 |1+ 2a^{\dagger}a |\Psi_1\rangle = \langle\Psi_1 | 1+2b^{\dagger}b |\Psi_1\rangle\\
&=& \mathrm{tr} ( 1+ 2 \sum_{k=0}^3 a^{\dagger}a \; \lambda_k |\phi_k\rangle \langle \phi_k|)\\
&=&  1+ 2 \sum_{k=0}^3 \lambda_k \langle \phi_k| a^{\dagger}a | \phi_k\rangle \\
&=& 1 + 2 \alpha^2 \sum_{k=0}^3 \lambda_k \frac{\lambda_{k-1}}{\lambda_k} \\
&=& 1 + 2 \alpha^2.
\end{eqnarray}

The correlation term $Z_1$ of the covariance matrix is given by
\begin{eqnarray}
Z_1 &=& \langle \Psi_1 |a b + a^{\dagger} b^{\dagger}|\Psi_1 \rangle\\
&=& 2 \mathcal{R}e \langle \Psi_1 |a b|\Psi_1 \rangle.
\end{eqnarray}
One has:
\begin{eqnarray}
a b |\Psi_1\rangle &=& ab \sum_{k=0}^3 \sqrt{\lambda_k} |\phi_k\rangle |\phi_k\rangle\\
&=& \alpha^2 \sum_{k=0}^3 \frac{\lambda_{k-1}}{\lambda_k} \sqrt{\lambda_k}  |\phi_{k-1}\rangle |\phi_{k-1}\rangle
\end{eqnarray}
where addition should be understood modulo 4. 
Finally, we obtain:
\begin{equation}
Z_1 = 2 \alpha^2 \sum_{k=0}^3 \frac{\lambda_{k-1}^{3/2}}{\lambda_k^{1/2}}.
\end{equation}

It may be noticed that at the lowest order in $\alpha$, the states $|\phi_k\rangle$ are simply the number states $|k\rangle$ for $k= 0$, 1, 2, 3, which are independent of $\alpha$. The states $|\psi_k\rangle$ are four orthogonal  linear combinations of these four number states, with coefficients of the form $e^{i p \pi/4}$, where $p$ is an integer. 

In addition, the state $| \Psi_1 \rangle$ is simply $(1-\alpha^2/2) |0 0 \rangle + \alpha^2 |1 1 \rangle$, which is also the lowest-order (Gaussian) EPR state. Correspondingly, $Z_1 = Z_{EPR} = 2 \alpha = \sqrt{2 V_A}$, in the limit where $\alpha$ tends to 0.

Since the entangled state is already Gaussian in this regime, no decoy states are needed, and Ref. \cite{LG09} can be used directly to establish the unconditional security of the protocol. Unfortunately, this approach is restricted to values of $\alpha$ which are too small to be useful in practice; this is why the more powerful proof presented in the present paper is needed.

\subsection{Eight-dimensional protocol: $d=8$}

The partial trace $\sigma_\mathrm{key}^8=\mathrm{tr}_A \left(|\Psi_8\rangle \!\langle \Psi_8|\right)$ is defined by the modulation scheme: it is the uniform mixture of quadrimodal coherent states over a real 7-dimensional sphere: 
\begin{equation}
\mathrm{tr}_A \left(|\Psi_8\rangle \!\langle \Psi_8|\right) :=\int_{\mathcal{S}_{\alpha}} |\alpha_1, \alpha_2, \alpha_3, \alpha_4\rangle \! \langle \alpha_1, \alpha_2, \alpha_3, \alpha_4| \d S
\end{equation}
where $ |\alpha_1, \alpha_2, \alpha_3, \alpha_4\rangle  := |\alpha_1\rangle | \alpha_2\rangle | \alpha_3\rangle | \alpha_4\rangle$ and the sphere $\mathcal{S}_{\alpha}$ is defined as
\begin{multline}
\mathcal{S}_{\alpha} \equiv \{ (\alpha_1,\alpha_2,\alpha_3,\alpha_4) \in \mathbb{C}^4 \\
\mathrm{s.t.} |\alpha_{4k}|^2+|\alpha_{4k+1}|^2+|\alpha_{4k+2}|^2+|\alpha_{4k+3}|^2 = 4 \alpha^2 \},
\end{multline}
and $\d S$ is the Haar measure on $\mathcal{S}_{\alpha}$.
Because this state is a four-mode orthogonally invariant state (by construction), it can be written as \cite{LC09}:
\begin{equation}
\mathrm{tr}_A \left(|\Psi_8\rangle \!\langle \Psi_8|\right) = \sum_{k=0}^{\infty} \lambda_k \, \sigma_k^4, 
\end{equation}
where 
\begin{equation}
\sigma_k^4 = \frac{1}{{k+3 \choose 3}} \sum_{\stackrel{k_1 \cdots k_4}{\;\;\mathrm {s.t.}\; \sum_i k_i = k }} |k_1,k_2,k_3,k_4\rangle\!\langle k_1,k_2,k_3,k_4|.
\end{equation}
In order to determine the $\{\lambda_k\}_{k=0, \cdots,\infty}$, we compute the probability $\text{Pr}(k)$ of finding $k$ photons in the four-mode state $\mathrm{tr}_A \left(|\Psi_8\rangle \!\langle \Psi_8|\right)$:
\begin{eqnarray}
\text{Pr}(k) &=& \text{tr} (\mathrm{tr}_A \left(|\Psi_8\rangle \!\langle \Psi_8|\right) \sigma_k^4)\\
&=& \langle 2\alpha|\langle 0|\langle 0|\langle 0| \, \sigma_k^4 \, |2\alpha \rangle |0\rangle|0\rangle|0\rangle,
\end{eqnarray}
since $|2\alpha \rangle |0\rangle|0\rangle|0\rangle \in \mathcal{S}_4$.
Because the coherent state $|0\rangle$, which refers to the vacuum, does not contain any photon, one has:
\begin{eqnarray}
\text{Pr}(k) &=& \langle 2\alpha| \sigma_k^4 |2\alpha \rangle \\
&=& e^{-4 \alpha^2} \frac{(2\alpha)^{2k}}{k!}\\
&=& \lambda_k.
\end{eqnarray}
We therefore get the expression:
\begin{equation}
\mathrm{tr}_A \left(|\Psi_8\rangle \!\langle \Psi_8|\right) = e^{-4\alpha^2} \sum_{k=0}^{\infty} \, \frac{(2\alpha)^{2k}}{k!} \, \sigma_k^4.
\end{equation}

Finally $|\Psi_8\rangle $ is defined as:
\begin{equation}
|\Psi_8\rangle := e^{-2\alpha^2} \sum_{k=0}^{\infty} \frac{(2\alpha)^k}{\sqrt{k!}} \, |\psi_k^4\rangle,
\end{equation}
where
\begin{equation}
|\psi_k^4\rangle = \frac{1}{\sqrt{{k+3 \choose 3}}} \sum_{\stackrel{k_1 \cdots k_4}{\;\;\mathrm {s.t.}\; \sum_i k_i = k }} |k_1,k_2,k_3,k_4\rangle | k_1,k_2,k_3,k_4 \rangle.
\end{equation}

Let us compute the covariance matrix $\Gamma_8$ of $|\Psi_8\rangle$. It has the form 
\begin{equation}
\Gamma_8 = 
\bigoplus_{i=1}^4 \left(
\begin{array}{cc}
X \mathbbm{1}_2 & Z_8 \sigma_z\\
Z_8 \sigma_z & X \mathbbm{1}_2
\end{array}
\right).
\end{equation}
where 
\begin{eqnarray}
X &= & \langle \Psi_8| 1 + 2 a_1^{\dagger} a_1 | \Psi_8 \rangle = \langle \Psi_8| 1 + 2 b_1^{\dagger} b_1 | \Psi_8 \rangle\\
Z_8 &= & \langle \Psi_8| a_1 b_1 + a_1^{\dagger} b_1^{\dagger} | \Psi_8 \rangle
\end{eqnarray}
where $a_1$ and $b_1$ refer to Alice and Bob's annihilation operators relative to the first mode. 

Tracing $|\psi_k^4\rangle$ over the last three modes gives $\rho_k^1$:
\begin{equation}
\rho_k^1 = \frac{1}{{k+3 \choose 3}} \sum_{l=0}^k {k-l+2 \choose 2} |l,l\rangle \!\langle l,l|.
\end{equation}
One immediately has:
\begin{equation}
\langle \Psi_8 |a_1^{\dagger}a_1|\Psi_8\rangle = \frac{1}{{k+ 3 \choose 3}} \sum_{l=0}^k l {k-l+2 \choose 2} = \frac{k}{4}.
\end{equation}
Then,
\begin{eqnarray}
\text{tr} (a_1^{\dagger}a_1 \rho^4) &=& \sum_{k=0}^{\infty} e^{-4 \alpha^2} \frac{(2\alpha)^{2k}}{k!} \frac{k}{4}\\
&=& \alpha^2,
\end{eqnarray}
which gives $X =1 + 2\alpha^2$.\\
Let us now compute $Z_8 = \langle \Psi_8| a_1 b_1 + a_1^{\dagger} b_1^{\dagger} | \Psi_8 \rangle$. First, one notes that $\langle \phi_l^4 |a_1 b_1 | \psi_k^4\rangle = 0$ except if $l = k-1$. Some combinatorics shows that
\begin{eqnarray}
\label{Z_correlation_psi4}
\langle \phi_{k-1}^4 |a_1 b_1 | \psi_k^4\rangle &=& \frac{1}{\sqrt{{k+3 \choose 3} {k+2 \choose 3}}}  \sum_{l=0}^k l {k-l+2 \choose 2 }\\
&=& \frac{1}{4} \sqrt{k(k+3)}.
\end{eqnarray}
Using the expression of $|\Psi_8\rangle$, one obtains
\begin{equation}
\langle \Psi_8 |a_1 b_1 |\Psi_8\rangle = \frac{1}{4} e^{-4\alpha^2} \sum_{k=0}^{\infty} \frac{\sqrt{k+4}}{k!} (2\alpha)^{2k+1},
\end{equation}
and finally
\begin{equation}
Z_8 = \frac{1}{2} e^{-4\alpha^2} \sum_{k=0}^{\infty} \frac{\sqrt{k+4}}{k!} (2\alpha)^{2k+1}.
\end{equation}

\section{Symmetrization of the protocol}
\label{active}

A quantum key distribution protocol is described as a map $\mathcal{E}$ \cite{CKR09}:
\begin{equation}
\mathcal{E}: 
\rho_{AB} \longmapsto (\mathcal{S}_A, \mathcal{S}_B,\mathcal{C})
\end{equation}
where $\rho_{AB} \in (\mathcal{H}_A \otimes \mathcal{H}_B)^{\otimes n}$ is the $n$-mode bipartite state shared by Alice and Bob at the end of the distribution phase (in the entanglement-based protocol), $S_A$ and $S_B$ are respectively Alice and Bob's final keys and $\mathcal{C}$ is a transcript of all classical communication as well as Alice's and Bob's raw data.

A protocol $\mathcal{E}$ is said to be invariant under some set of transformations $\mathcal{G}$ if for any element $g \in \mathcal{G}$, there exists a CPTP map $\mathcal{K}_g$ such that:
\begin{equation}
\mathcal{E} \circ g = \mathcal{K}_g \circ \mathcal{E}. 
\end{equation}

Let us consider the uniform measure $\mu_\mathcal{G}$ on $\mathcal{G}$. If the protocol $\mathcal{E}$ is invariant under the set $\mathcal{G}$, then it is sufficient to prove the security of $\mathcal{E}$ for states displaying the same symmetry. In particular, it is sufficient to consider states of the form:
\begin{equation}
\bar{\rho}_{AB} = \frac{1}{\mu_\mathcal{G}(\mathcal{G})} \int_\mathcal{G} \mathrm{d} \mu_\mathcal{G}(g) \, g(\rho_{AB}),
\end{equation}
for any $\rho_{AB} \in (\mathcal{H}_A \otimes \mathcal{H}_B)^{\otimes n}$ and where $g(\rho_{AB})$ is the image of the mode state $\rho_{AB}$ by $g$.

If we consider here for $\mathcal{G}$ the group of conjugate passive symplectic operations applied on Alice's $n$ modes and Bob's $n$ modes (in phase space, such operations are simply conjugate orthogonal transformations), then for a given operation $g$ applied to the state, the map $\mathcal{K}_g$ is obtained by applying the orthogonal transformations corresponding to $g$ on the classical data measured by Alice and Bob. If the protocol is invariant under this whole group, then it is sufficient to look at Gaussian states to prove security against collective attacks \cite{LG10b}. 
In the prepare and measure version of the protocol, this group becomes the orthogonal group $O(2n)$.

One simple way to ensure that a protocol is indeed invariant under a set $\mathcal{G}$ of transformations is for Alice and Bob to actively apply random transformations of $\mathcal{G}$ to their states.

In particular, if Alice and Bob both apply random orthogonal transformations to their classical vectors in the prepare and measure protocol, then the security analysis can be done assuming that they share a Gaussian state in the entangled version of the protocol.

For our security proof, the goal of the symmetrization is to make sure that the state shared by Alice and Bob is as isotropic as possible. Indeed, remember that we need to estimate the covariance matrix of the state shared by Alice and Bob after Alice's generalized measurement $\{\Pi_d, \mathbbm{1}-\Pi_d\}$. The only problem that could potentially happen would be that Eve guesses which states might be used for key distillation and which ones might be used for parameter estimation and that she manages to somehow play with Alice and Bob's correlations in order to fool them into overestimating their correlations. To totally prevent such a (quite unrealistic) scenario, it is sufficient to symmetrize the state so that there are no privileged directions in phase space that Eve could exploit.
Hence, in practice, the symmetrization does not require to apply random conjugate passive symplectic operations chosen with the uniform measure over the whole set of such operations: a smaller subset should be efficient. A very conservative quantitative criterium to evaluate the quality of such a set would be for instance the distance between the partial trace of $\bar{\rho}_{AB}$ once we trace out $n-1$ modes and the Gaussian state with the same first two moments.

\subsection{Active symmetrization}

To make sure that the protocol is indeed invariant under specific transformations, we apply an active symmetrization step to the state $\rho_{AB}$ before applying the protocol. 
The transformations we consider are conjugate passive symplectic operations applied on Alice's $n$ modes and Bob's $n$ modes, which therefore correspond to orthogonal transformations applied to their classical vectors in the prepare and measure protocol. For simplicity, we restrict the discussion to this prepare and measure scheme in the following.

The active symmetrization requires us to choose a subset $\mathcal{F}$ of the orthogonal group and for Alice and Bob, and to apply the same element $f \in \mathcal{F}$ (chosen uniformly at random) to their data before starting the postprocessing.
As we stressed above, taking for $\mathcal{F}$ the whole orthogonal group is not necessary in practice.
Hence, we want $\mathcal{F}$ to be a subset of the orthogonal group with the following properties: drawing a random element $f$ from the uniform measure on $\mathcal{F}$ should be doable with resources (time and alea generation) scaling at most linearily in $n$, the description of $f$ should also be at most linear in $n$ and applying $f$ (or $f^{-1}$) to a random vector of $\mathbbm{R}^n$ should also be at most linear in $n$. These conditions ensure that the protocol with the active symmetrization remains practical.
Moreover, the symmetrization should work as well as possible, meaning that $\mathcal{F}$ should symmetrize the state as much as possible.

We give examples of such possible subsets $\mathcal{F}$ in the next subsection.

\subsection{Construction of practical symmetrizations}

Let us describe a recursive algorithm that allows one to draw an orthogonal transformation with the Haar measure on $O(n)$.

If we assume that we already drew a random transformation $\tilde{R}_{n-1}$ from the Haar measure on $O(n-1)$, then let us note $R_{n-1}=\mathbbm{1} \oplus \tilde{R}_{n-1}$ the orthogonal transformation in $O(n)$ acting as the identity on the first element of the canonical basis of $\mathbbm{R}^n$ and as $\tilde{R}_{n-1}$ on the last $n-1$ elements of the basis.

Then, let us draw uniformly at random a unit vector $u_n$ on the sphere $\mathcal{S}^{n-1}$ and define the following Householder reflection $H_n$:
\begin{equation}
H_n = \mathbbm{1}_n - 2 u_n u_n^T.
\end{equation}
Note that drawing $u_n$ can be done in linear time simply by drawing $n$ random normal variables and normalizing the obtained vector. Also, applying $H_n$ to any vector can be done in linear time in $n$.

Finally, one can show that random orthogonal transformation $R_n = H_n R_{n-1}$ follows the Haar distribution of $O(n)$ \cite{mez06}. In particular, one has:
\begin{equation}
R_n = \prod_{k=n}^1 \tilde{H}_k
\end{equation}
where one defines $\tilde{H}_k = \mathbbm{1}_{n-k}\oplus H_k$.

Hence, drawing, describing and applying a random orthogonal transformation from $O(n)$ are tasks with complexity quadratic in $n$. 
We define the subset $\mathcal{F}_k$ of the orthogonal group as corresponding to the set of compositions of $k$ such Householder reflections, that is the last $k$ steps of the algorithm described above. For instance, $\mathcal{F}_1$ corresponds to the set of Householder reflections with respect to a hyperplane of $\mathbb{R}^n$, and $\mathcal{F}_n$ is the orthogonal group $O(n)$.
One can also define a family of measures $\mu_1, \mu_2, \cdots, \mu_n$ on $O(n)$ corresponding to the uniform measures of $\mathcal{F}_1, \mathcal{F}_2, \cdots, \mathcal{F}_n$.

A complete symmetrization would imply performing orthogonal transformations on both Alice and Bob's data chosen randomly with the measure $\mu_n$, but for all practical purposes, it seems that $\mu_1$ already provides a high level of symmetrization.

\section{Full protocol with the symmetrization step}
\label{symmetrized_protocol}

We present here two different schemes, depending on the choice of modulation which can be either fully Gaussian or consists of a non-Gaussian modulation supplemented by appropriate decoy states. The former modulation is compatible with both a homodyne or a heterodyne detection and corresponds to the technique detailed in Appendix \ref{noisy} while the latter, which is more efficient in terms of resources, is only compatible with a heterodyne detection (and is detailed in Appendix \ref{optimization}).
These schemes include the active symmetrization introduced in Appendix \ref{active}.

\subsection{Fully Gaussian modulation}  

The full protocol is the following:
\begin{itemize}
\item Alice draws $2n$ random variables $x_1, x_2, \cdots, x_{2n}$ from a centered normal distribution with the appropriate variance. These form a vector $x \in \mathbbm{R}^{2n}$.
\item Alice sends the states $|\alpha_1\rangle, \cdots, |\alpha_k\rangle, \cdots, |\alpha_{n}\rangle$ to Bob, with $|\alpha_k\rangle = |x_{2k} + i x_{2k+1}\rangle$.
\item Bob receives the states after the quantum channel and measures them, with either a homodyne detection or a heterodyne detection. In the case of a heterodyne detection, he obtains a $2n$-dimensional vector $y$. In the case of a homodyne detection, he obtains an $n$-dimensional vector $y$, then informs Alice about his choices of measured quadratures ($x$ or $p$ for each state); Alice only keeps the relevant coordinates in her data in order to form a new $n$-dimensional vector $x$.
\item Alice randomly draws a random orthogonal transformation $R$ from the orthogonal group $O(2n)$ (or $O(n)$ for a homodyne detection). In theory, to achieve a perfect symmetrization of the state, Alice should draw $R$ with the Haar measure on $O(2n)$. However, in practice, $R$ can be chosen uniformly in a well-chosen subset of $O(2n)$ which has the advantage of allowing for efficient descriptions of its elements, such as one of the measures $\mu_k$ defined in Appendix \ref{active}.
\item Alice describes $R$ to Bob through the classical communication channel, and both parties apply $R$ to their respective vector, hence obtaining $x'=Rx$ and $y'=Ry$.
\item Alice chooses randomly $n_\mathrm{PE}$ coordinates that are used for parameter estimation.
\item The next step is where lies the novelty of our protocols. For instance, in the so-called four-state protocol, Alice considers the coordinates $x_k'$ which were not used for parameter estimation and keeps only the ones such that $|x_k'|$ is sufficiently close to a predetermined value. In the case of the eight-dimensional protocol (with heterodyne detection for instance), Alice divides her data into blocks of size 8 and keeps the blocks for which the euclidean norm is close to a predetermined value (see Appendix \ref{noisy} for details). Alice informs Bob of the indices that she keeps. The rest of the data are discarded. At this point, Alice and Bob have classical correlations for which an efficient reconciliation protocol is available (see Appendix \ref{reduction}).
\end{itemize}

\subsection{Non-Gaussian modulation and decoy states}

In this scheme, the positions of the states used for parameter estimation are chosen randomly \emph{beforehand} by Alice. Let us consider for simplicity the case of the eight-dimensional protocol.
\begin{itemize}
\item Alice draws $n$ 8-dimensional random vectors, each chosen from one of the three following distributions: random vectors on the 7-dimensional sphere with the appropriate radius (these data correspond to the non-Gaussian modulation which will be used for the key distillation), random vectors on the 7-dimensional sphere with an appropriately fluctuating radius (these are the decoy states which will be discarded at the end of the protocol: the mixture of these states with the previous ones should be indistinguishable from a true Gaussian distribution) or Gaussian vectors which are used for parameter estimation. Alice hence obtains an $8n$-dimensional vector $x$ for which each subset of length 8 corresponds either to legitimate information, decoy data that will be discarded or data used for parameter estimation.
\item Alice randomly draws a random orthogonal transformation $R$ from the orthogonal group $O(8n)$. In theory, to achieve a perfect symmetrization of the state, Alice should draw $R$ with the Haar measure on $O(8n)$. However, in practice, $R$ can be chosen uniformly in a well-chosen subset of $O(2n)$ which has the advantage of allowing for efficient descriptions of its elements, such as one of the measures $\mu_k$ defined in Appendix \ref{active}.
\item Alice computes the vector $x' = Rx$, which is the image of $x$ by the orthogonal transformation $R$ and uses this vector for her modulation. Hence she sends the states $|\alpha_1\rangle, \cdots, |\alpha_k\rangle, \cdots, |\alpha_{4n}\rangle$ to Bob, with $|\alpha_k\rangle = |x_{2k}' + i x_{2k+1}'\rangle$.
\item Bob receives the states after the quantum channel and measures them, with a heterodyne detection. He obtains an $8n$-dimensional vector $y'$. 
\item Alice describes $R$ to Bob through the classical communication channel. Bob applies $R^{-1}$ to his vector $y'$ and obtains $y= R^{-1}y'$.
\item Alice reveals which subsets of length eight should be kept for the key distillation, which ones should be discarded (as they correspond to decoy states) and which ones should be used for parameter estimation.
\item At this point, Alice and Bob have classical correlations for which an efficient reconciliation protocol is available (see Appendix \ref{reduction}).
\end{itemize}

\end{document}